\documentclass[preprint2]{aastex}
\usepackage{times}
\newcommand{\kms}{km s$^{-1}$}
\newcommand{\acm}{$^{\prime}$}

\newcommand{\obc}{$\rm \left[ OBC97 \right]$ }

\newcommand{\degree}{$^\circ$}
\newcommand{\HI}{\mbox{H\,{\sc i}}}
\input{epsf}
\slugcomment{Accepted by AJ}
\shortauthors{O'Neil, et al.}

\shorttitle{New Reference Galaxy Standards for \HI\ Emission Observations}
\begin{document}
\title{New Reference Galaxy Standards for \HI\ Emission Observations}
\author{K. O'Neil\altaffilmark{1}}
\affil{NRAO; PO Box 2; Green Bank, WV 24944}
\altaffiltext{1}{Work done while at Arecibo Observatory}
\email{koneil@nrao.edu}

\begin{abstract}
We have taken advantage of the improved baselines and higher sensitivity
available with the upgraded Arecibo 305m telescope to create a new \HI\ spectral
line catalog of disk galaxies which can be used as a reference catalog for anyone interested
in 21-cm spectral line work.
In all 108 galaxies were observed, covering 24h of the sky
at declinations between 0$^\circ < \delta <36^\circ$ and velocities between
0 -- 25,000 \kms.  The majority of the galaxies were  observed at least two times on different nights
to avoid problems with radio frequency intereference, 
baselines fluctuations, etc.  Comparing our measured values with all those
available in the literature show that while large individual variations may exist,
the average differences between the measurements to be zero.
In all we have considerable confidence in our measurements, and the
resultant catalog should be  extremely useful as a well defined reference catalog for anyone interested
in 21-cm spectral line work.
\end{abstract}


\section{Introduction}

The 21-cm neutral hydrogen line is possibly the most commonly observed spectral line by
centimeter wavelength telescopes.  Yet a brief perusal through any collection of
of \HI\ observations on a gas-rich galaxy will often find
flux measurements varying by 5\% -- 100\%, or more (Huchtmeier \& Richter 1989;
NED\footnote{NED, the NASA/IPAC Extragalactic Database,
is operated by the Jet Propulsion Laboratory, California Institute of Technology,
under contract with the National Aeronautics and Space Administration.}).
The reasons behind these discrepancies are many-fold.  The simplest explanation for
the differences is various telescopes' beam sizes and shapes.  That is, observing a
galaxy with \HI\ spread over a 10$^\prime$ region 
with the Arecibo 305m telescope (3.6$^\prime$ FWHM beam at
1.4 GHz) would certainly result in a different flux measurement than if that same
object were observed with the Green Bank Telescope (9.2$^\prime$ FWHM beam at 1.4 GHz).
Yet while the differences in beam sizes and patterns may account for some of the discrepancies seen,
it cannot be the only explanation, as only a fraction of the known galaxies
have diameters larger than the Arecibo beam.  (Roughly 15\% of the galaxies in the UGC catalog
have optical diameters $>$2$^\prime$  -- Nilson 1973.)
Instead, or in addition, to the previous explanation, the discrepancies in the different measurements
may be due to inaccuracies in the hardware and software used for data taking and analysis.
These can include pointing offsets, unstable hardware resulting
in baseline variations, rapid (compared to the frequency of measurements) temperature fluctuations,
poor telescope focus, and over- or under-estimates of line flux due to inaccurate baseline fitting.
In addition, line flux measurements can suffer from erroneous values of both the
astronomical (gain) and system (temperature) calibration values due to
either inaccuracies in the measurements or
or measurement which are sampled too sparsely in time, frequency, or position
to accurately reflect the system.

As a result of the variations found in the different \HI\ galaxy catalogs, astronomers
wishing to obtain accurate flux values for a given galaxy are often forced to
average the various measurements found in published catalogs, invoke a selection effect
on the previous measurements, assuming some measurements are more accurate than others,
or simply re-observe the objects of interest with the hope that their data will be be more
reliable, on average, than that found in the literature.  While all of the
above methods are sound and should be encouraged, they can also be impractical, particularly
when the researcher is interested in comparing their own results against a consistent,
well-defined catalog with readily understood errors.  This is not meant to imply that
catalogs of \HI\ measurements found in the literature are in general unreliable,
simply that as the purpose of the majority of catalogs is to look for statistical
information regarding the neutral hydrogen content in the studied galaxies, the errors
associated with the measurements for an individual galaxy can be fairly high.

Previous efforts have been made to generate extremely accurate lists of \HI\ in galaxies
for calibration and references purposes (e.g. Tifft 1992; Schneider, et al. 1992;
Huchtmeier \& Richter 1988; Baiesi-Pillastrini \& Palumbo 1986).  However,
as a result of its recent upgrade \citep{salter02}, the Arecibo 305m telescope now offers
unprecedented stable baselines, increased sensitivity, and extremely good pointing.
As a consequence, we have taken advantage of Arecibo's exceptional performance
and have observed over 100 galaxies with the aim of offering a catalog of \HI\ observations
which can be used as a  reference catalog for anyone interested
in 21-cm spectral line work.  All of the data and results for this project are publicly available
online at both \url{http://www.gb.nrao.edu/$\sim$koneil/HIsurvey} and
\url{http://www.naic.edu/$\sim$astro/HIsurvey}.

\section{Catalog Selection}

\begin{figure}
\plotone{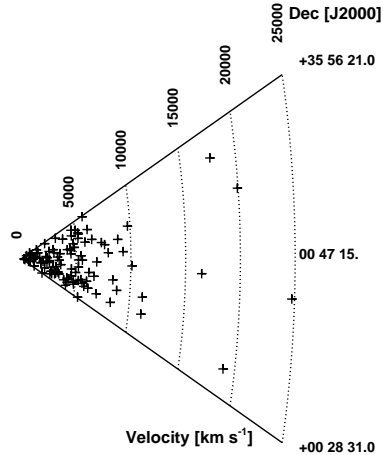}
\caption{Distribution of observed galaxies in Dec.-velocity space
\label{fig:gal_dist}}
\end{figure}

\begin{figure}
\plotone{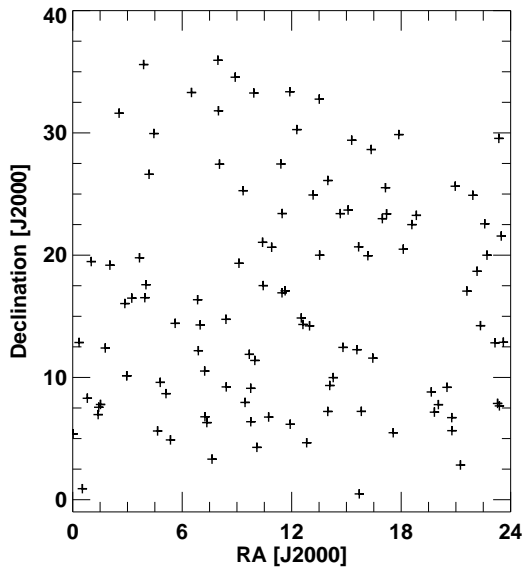}
\caption{Distribution of observed galaxies in RA-Dec space.
\label{fig:gal_dist2}}
\end{figure}

The galaxies chosen cover the complete Arecibo sky
($-1^\circ < \delta < 38.5^\circ$), with at
least three galaxies within any given hour of LST.
Additionally, an attempt was made to cover velocity
space from 0 -- 25,000 km s$^{-1}$.  However, our requirement that the
objects' signal-to-noise be reasonably high within a few minutes of observation (detailed below)
resulted in the majority of the objects lying at $<$8,000 \kms.  To remedy
this problem, a few objects were observed which do not meet the signal-to-noise
criteria.  These objects are noted in Table~\ref{tab:gals}, which also gives a
complete listing of galaxies observed and their coordinates, and Figures~\ref{fig:gal_dist}
and \ref{fig:gal_dist2}
show the distribution of the galaxies in velocity, Declination, and Right Ascension
space.

In addition to the above constraints, the selection for the catalog
was based on five major criteria.  The observed galaxies should:
\begin{itemize}
\item Have previous 21-cm line measurements indicating an emission
of at least 10 mJy. This ensures a signal-to-noise ratio at
Arecibo of $>$8 after five minutes of on-source observation (using
8.5 km s$^{-1}$ channel$^{-1}$ resolution); \item Have an optical
diameter, D$_{25}\;\le$ 2\acm\ to avoid the galaxy's \HI\
from extending past the $\sim$3.5\acm\ Arecibo beam; \item Be
isolated to a 10\acm, 1000 km s$^{-1}$ radius to avoid
contamination through the beam side-lobes; \item Have no
significant continuum emission.  This avoids residual standing
waves which arise from Arecibo's partially blocked aperture
\citep{ghosh01}.
\end{itemize}

No information regarding the distribution of the \HI\ for the observed 
galaxies was used in determining the catalog, in large part due to the 
fact that such information is not available for many of the galaxies observed.  
As a result, any discontinuities or asymmetries in the galaxies' 
\HI\ distribution was not accounted by this project.

\section{Observations, Calibration, and Data Reduction}

Data was taken with the L-narrow receiver between Aug, 2001 - July, 2002. Each galaxy was
observed with two polarization channels at two different resolutions (12.5 MHz/2048 channels for
1.3  km s$^{-1}$ channel$^{-1}$ resolution and 6.25 MHz/2048 channels for
0.63 km s$^{-1}$ channel$^{-1}$ resolution at 1420 MHz).
Standard position switching techniques were used for
all observations, tracking the same azimuth and zenith angle for the off-source observation as for
the on-source observation to reduce differences in the baseline shapes.
Observations were typically 5 minutes on-source and 5-minutes off-source, although
the total time was occasionally reduced if the telescope scheduling required it.
Temperature measurements (with a calibrated noise diode) were made after
each on+off observation pair.  Details on the noise diode calibration can be found on
the Arecibo web pages at \url{http://www.naic.edu/$\sim$astro/Lnarrow}.

A gain curve for the telescope was obtained through reducing all observations taken of
standard continuum calibrators, by any project, during the observing period.  A complete
description of the procedures used to determine the gain curve in this manner can be
found in \citet{Heiles01}.  Additionally, observations were made of standard continuum
calibrators every 2-3 hours during the project observations, with the results checked
against the determined telescope gain, to insure no anomalous behavior occurred in the
hardware during observations.

The variation between any individual measurement of a calibration source and the
gain measurement could range as high as 20\%, particularly in Arecibo's outer
declination reaches (32$^\circ \le \delta \le$ 36$^\circ$).  Smoothing the
individual measurements onto a two-dimensional fit reduces the overall
calibration error to less then 10\%.  To further reduce the error, sources were observed
at angles between 2$^\circ$ -- 14$^\circ$ from the zenith whenever possible to avoid the
regions where the residuals of the gain curve fit were high.  Finally, each object was observed
a number of different times, at different LSTs, so that any variations in the determined
two-dimensional gain curve are minimized.  The 
the end result is flux measurements which should be reliable at the 5\% level, or better.

To reduce the data, first the individual scans for each observation were averaged, and the
on- and off-source observations were combined.  The correct value for the gain curve was obtained
using the average azimuth and zenith from the individual one second data dumps.  The
gain and system temperature corrections were made after combining the on- and off-source
observations.  At this point baselines were subtracted (using first order polynomials)
and the flux, heliocentric velocity, and velocity widths were determined
from each individual observation.  This allowed for the determination of the potential
ranges of values which an astronomer might see when observing a given object.
Finally, the individual observations (without baseline subtractions) were combined to
give the highest possible signal-to-noise for a given galaxy.  Baselines were
subtracted and the  flux, heliocentric velocity, and velocity widths were determined
from this averaged observation.  In each case baseline fitting was done using a
minimum of 200 frequency channels on each side of the line of interest.
In sum, the observed, reduced data was
\[\rm Spectra\;=\;{{\langle On_i \rangle \;-\;\langle Off_i \rangle}\over{\langle Off_i \rangle}}
\; \times \;{{T_{sys}}\over{Gain}}\]
\[\rm Final\;Spectra\;=\;\langle Spectra_i\rangle\]

All baselines appeared to be best fit (in the $\chi^2$ sense) by first
order polynomials.  Any discrepancies between the baseline fits and the data 
are reflected in the r.m.s. measurements obtained.

\section{Final Results and Error Calculations \label{sec:final_res}}

The final results and associated errors are reported in Table~\ref{tab:gals}, and
the 8.5 \kms\ resolution spectra are in Figure~\ref{fig:HI_profs}.
(The 1.3 and 0.65 \kms\ resolution spectra are online at
\url{http://www.gb.nrao.edu/$\sim$koneil/HIsurvey}.)
Each spectra was analyzed three times, at three different resolutions --
0.65 \kms, 1.3 \kms, and 8.5 \kms, with the 8.5 \kms\ resolution 
data obtained through boxcar averaging the
0.65 \kms\ resolution data.
The baseline fit region (described in the last section) was also used for
determining the r.m.s. of the spectra.  The flux was found by fitting (by eye)
the outer edges of the galaxy, and all flux within the defined regions was summed.
The velocity widths were determined at 20\% and 50\% of the mean value of the

\onecolumn
\begin{figure*}
\plotone{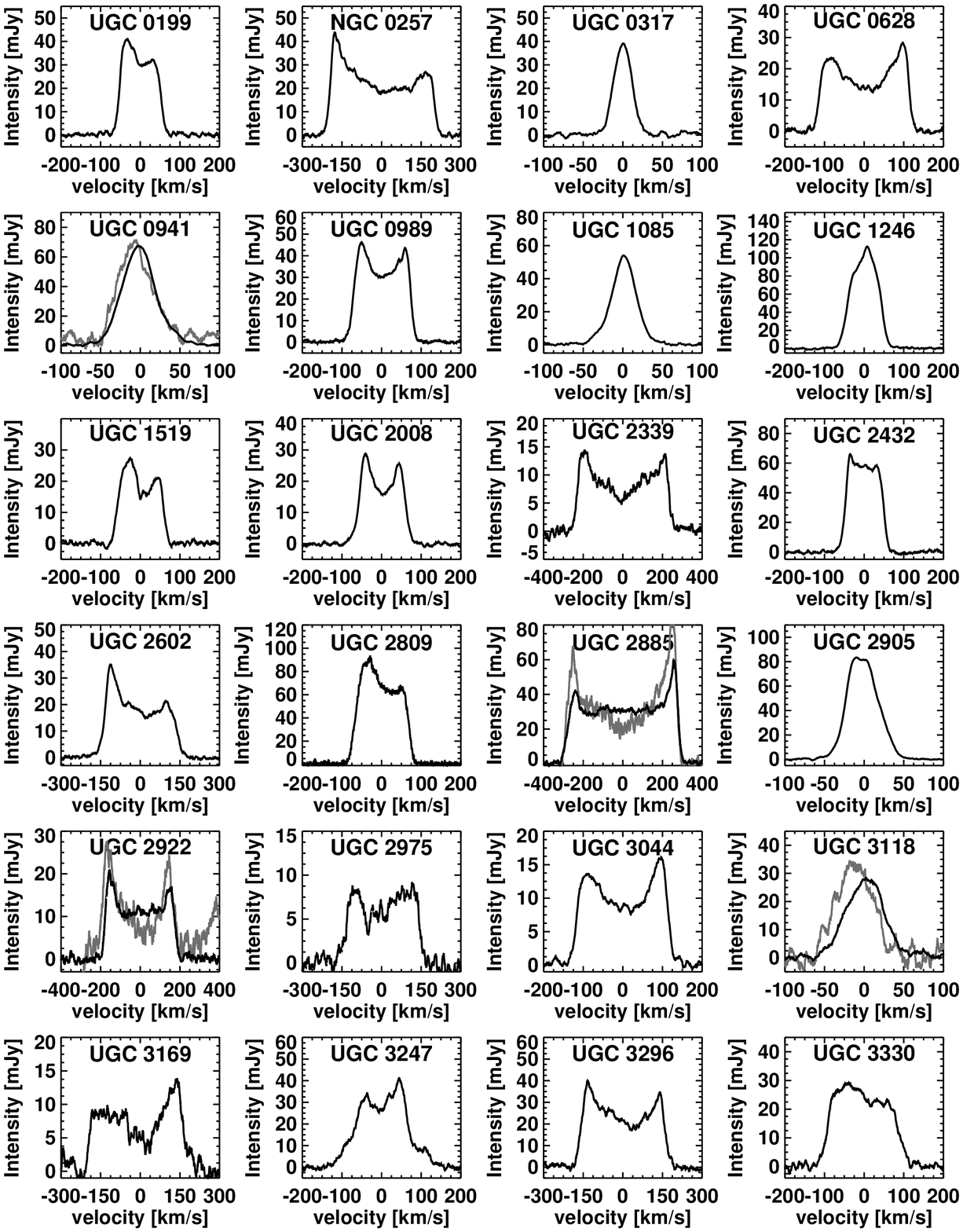}
\caption{8.5 \kms\ resolution spectra of the observed galaxies.  The black
lines are the Arecibo data and the gray lines are the GBT data.\label{fig:HI_profs}}
\end{figure*}

\addtocounter{figure}{-1}
\begin{figure*}
\plotone{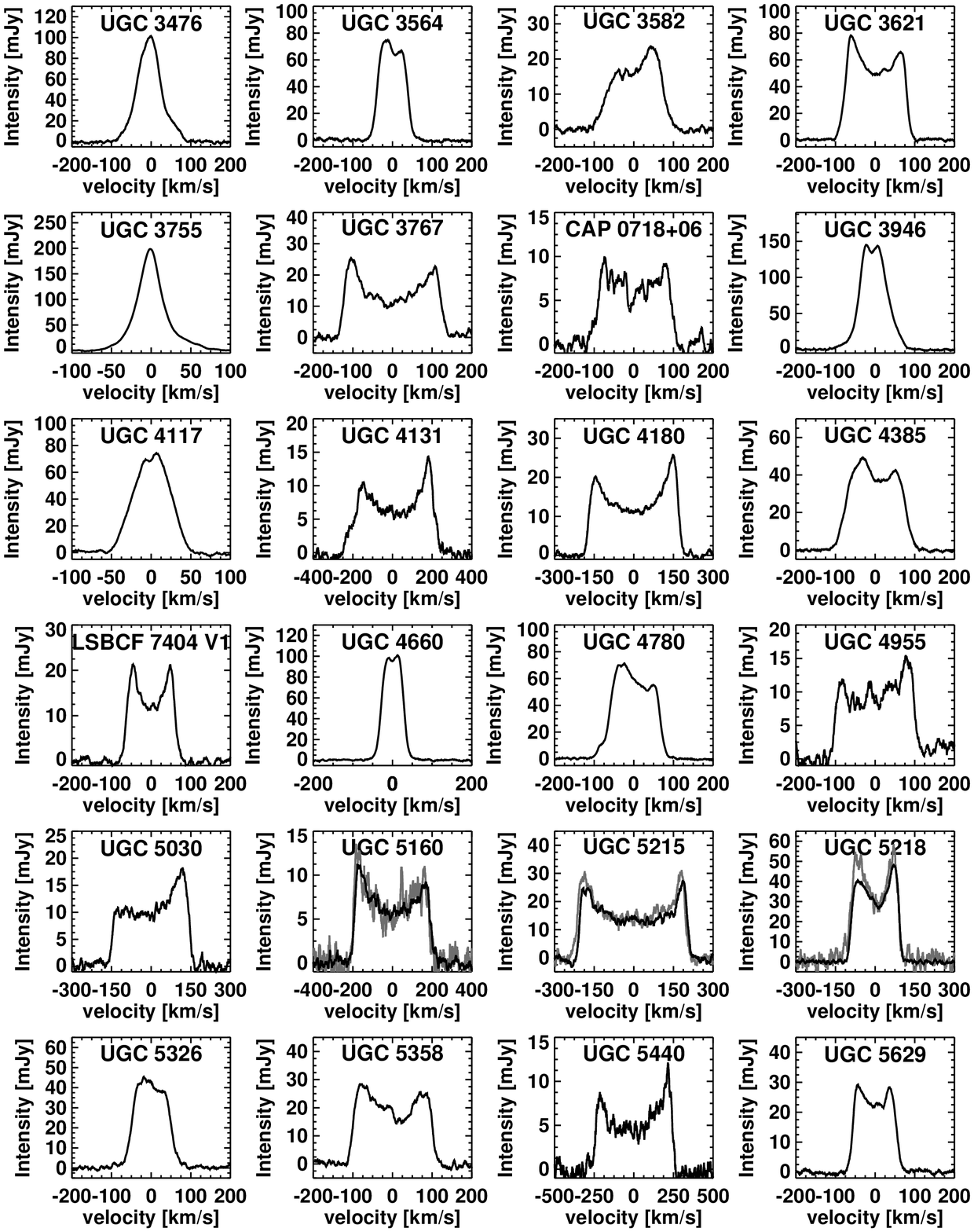}
\caption{\it cont.}
\end{figure*}

\addtocounter{figure}{-1}
\begin{figure*}
\plotone{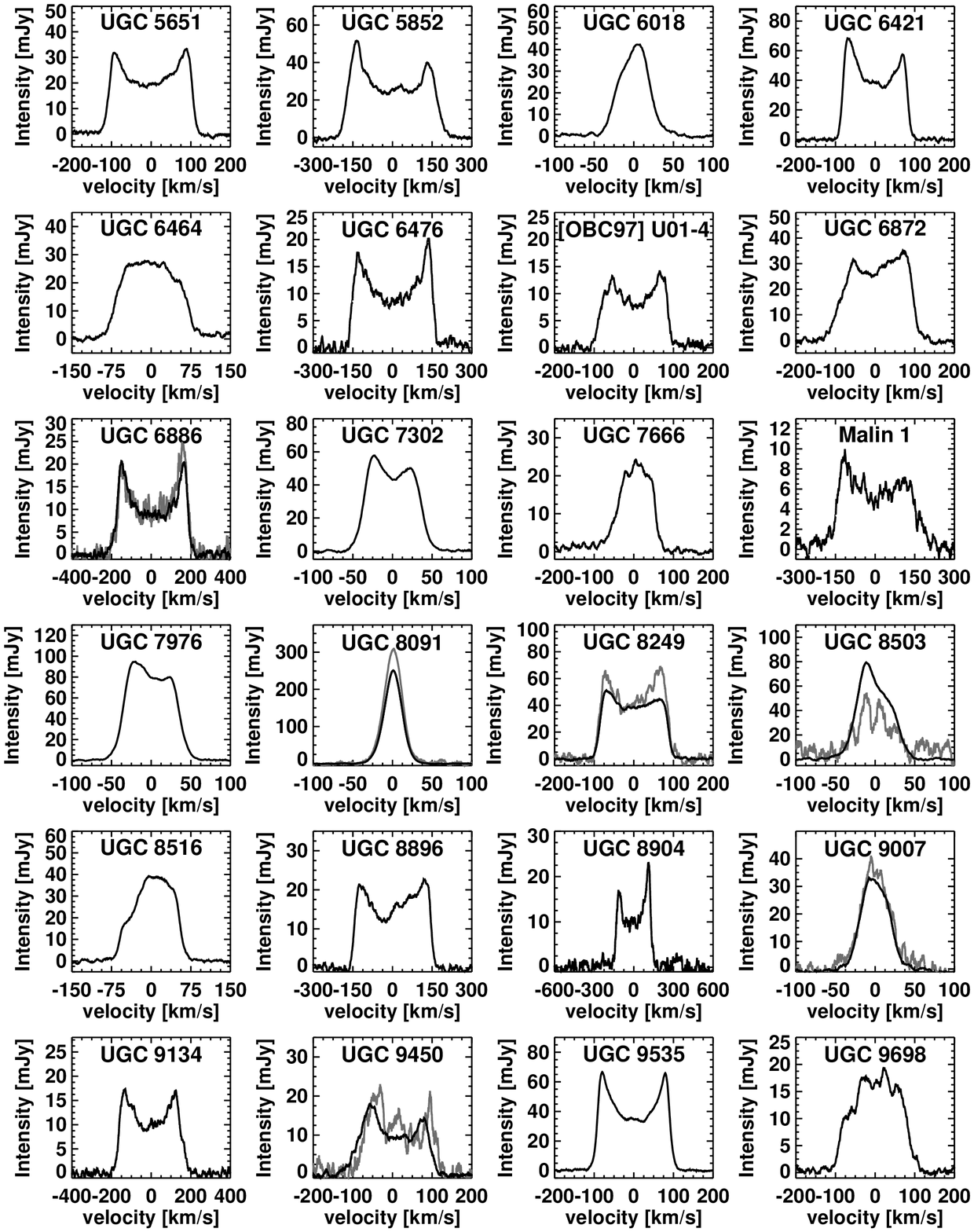}
\caption{\it cont.}
\end{figure*}

\addtocounter{figure}{-1}
\begin{figure*}
\plotone{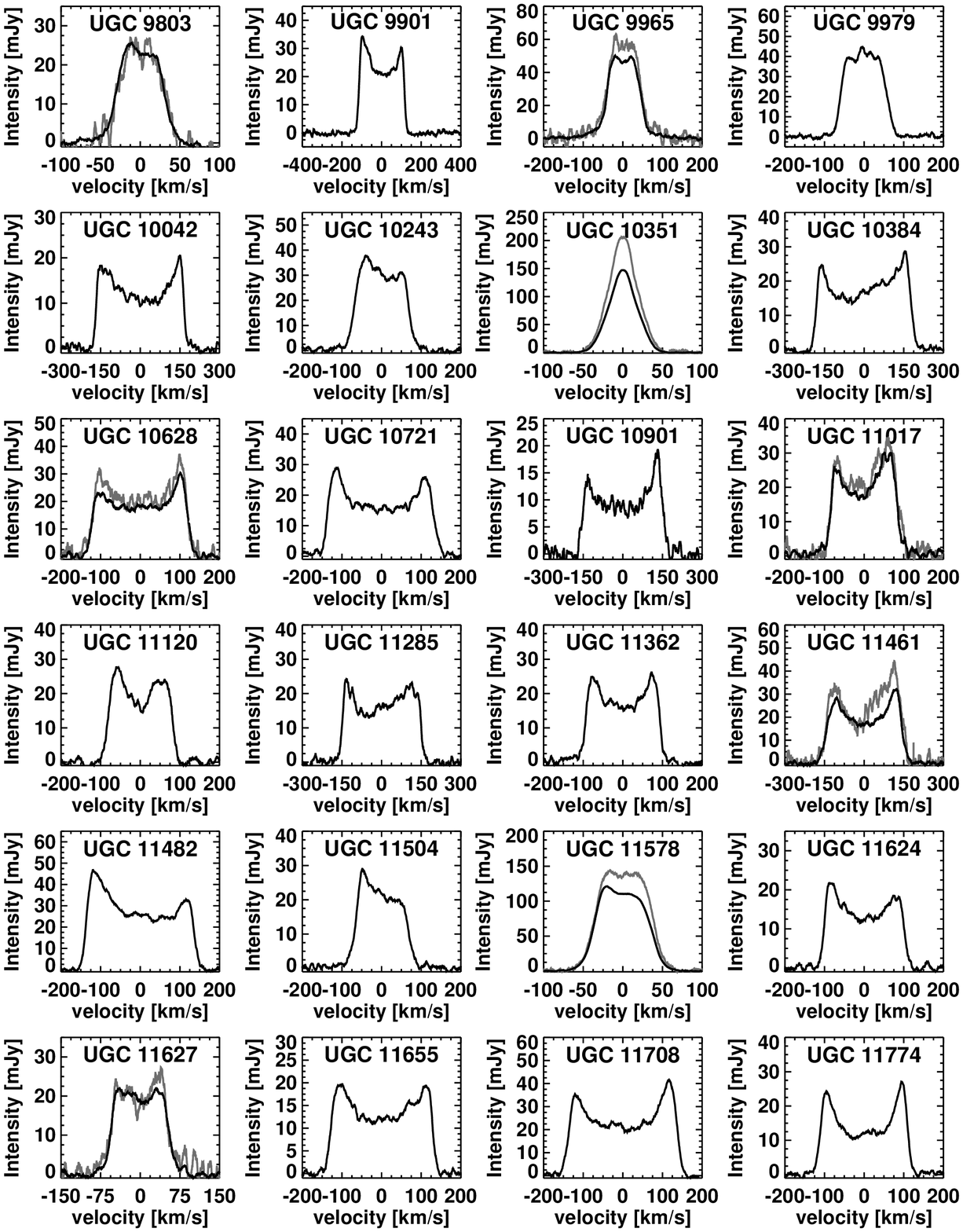}
\caption{\it cont.}
\end{figure*}

\addtocounter{figure}{-1}
\begin{figure*}
\plotone{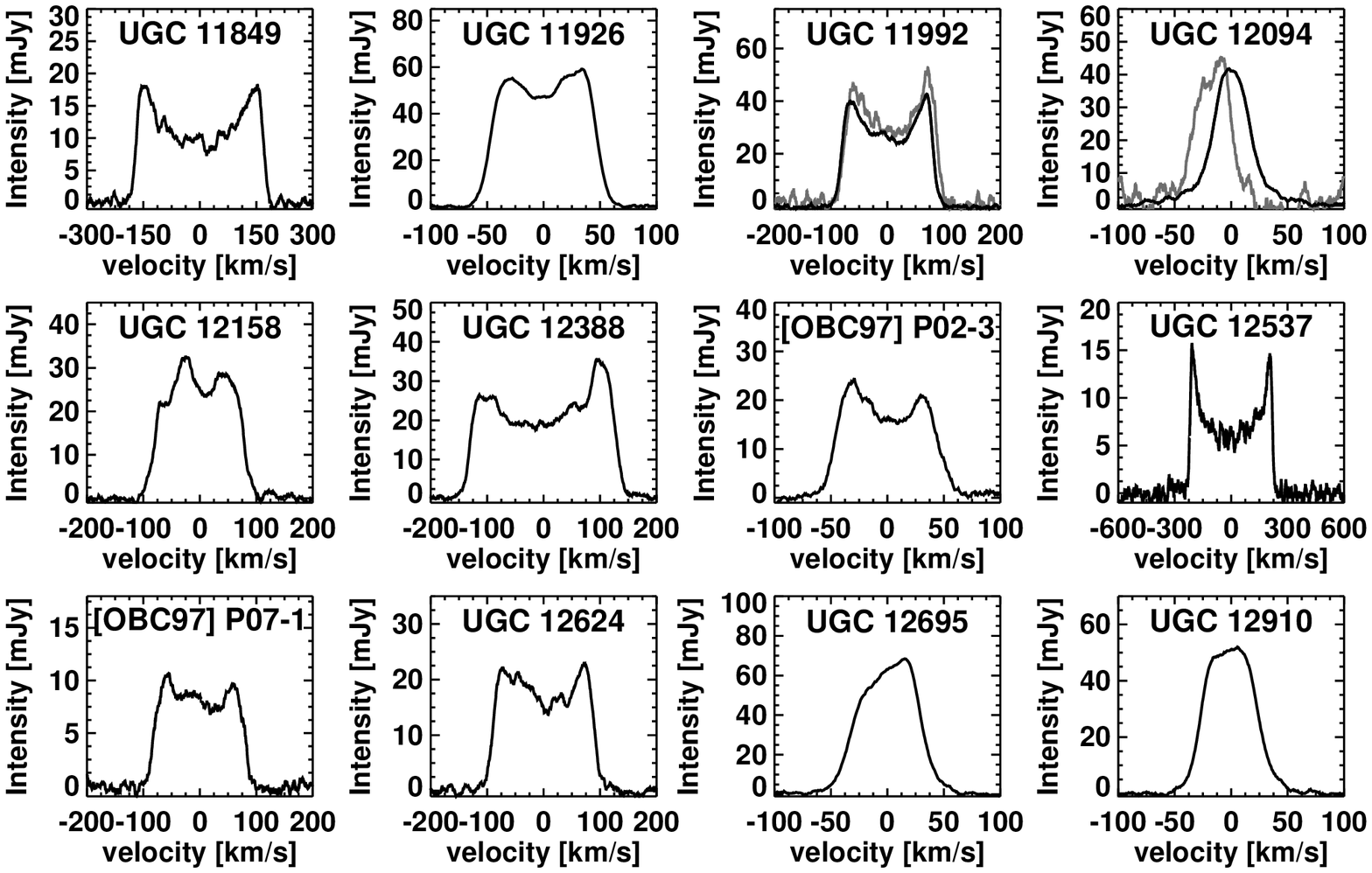}
\caption{\it cont.}
\end{figure*}
\twocolumn
\noindent selected region. Finally, the heliocentric velocity is equal to
$1/2\; \times\;\left(v_{high}\;-\;v_{low}\right)$, where $v_{low}$ and $v_{high}$
are equal to the low and high values of the heliocentric velocity at 
20\% or 50\% times mean value of two \HI\ peaks.

The results from the data analysis were then combined, and errors determined, as
described below.  For each galaxy both the individual observations and the average of
the individual observations were analyzed.  The results from the averaged observations
provide the most accurate measurements and are reported in Table~\ref{tab:gals}.  The
individual measurements were used only to determine the range of values found, and
are reported online
at http://www.gb.nrao.edu/$\sim$koneil/HIsurvey.

\subsection{Heliocentric Velocity}

The reported value for the heliocentric velocity comes from taking the mean value of
the three heliocentric velocity measurements obtained from the averaged (high signal-to-noise)
spectra.  To determine the error for this value, first the r.m.s. error at each velocity resolution
was determined through averaging the velocity measurements from the individual (not averaged)
observations.  The errors from the three velocity resolution were then added, in quadrature, to
determine the final reported error.

\subsection{Flux}

Fluxes were determined at each velocity resolution from the averaged (high signal-to-noise)
spectra of each galaxy.  The r.m.s. of the averaged spectra was found through examining
a minimum of 200 channels on either side of the spectral line of interest.  The observational
error was then determined as $\sigma_{flux, obs} = \sqrt{N} \times (channel\; width) \times \sigma_{rms}$,
where N is the number of channels (the velocity width divided by the channel width), and
$\sigma_{rms}$ is the rms error calculated.
The final error was determined through adding (in quadrature) the observational
error and a 5\% error assumed for the gain curve.
(See Section 3 for a description of the gain curve error).

\subsection{Velocity Widths}

Velocity widths were found at both 20\% and 50\% of the peak flux for each velocity resolution
using the averaged (high signal-to-noise) spectra of every galaxy.  The error reported
for the velocity widths are the r.m.s. errors from the average of the individual
(not averaged) observations.  On the occasion where the r.m.s. error was less than half the
channel resolution, the error was set equal to half the channel resolution.

\begin{figure}
\plotone{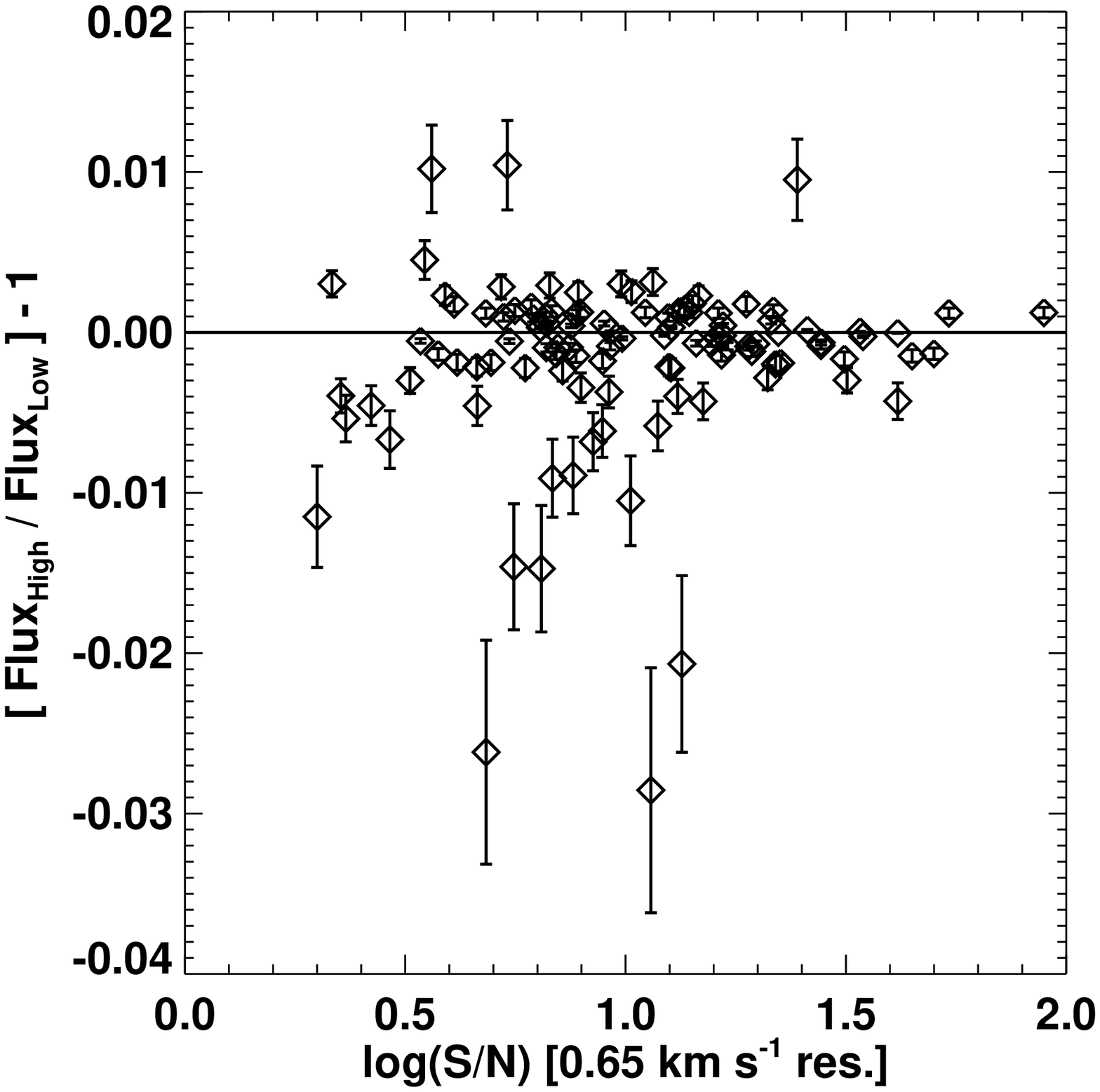}
\caption{Comparison of the  r.m.s. differences for the lowest
and highest resolution data measured against the average signal-to-noise of the spectral
line. While the average difference between the two resolutions (${{Flux_{high}}\over{Flux_{low}}}\;-\;1$)
is zero, a marked difference from zero for the r.m.s. can be seen for the lower
S/N.  (Average signal to noise was found through dividing the total flux by the r.m.s. noise
for each galaxy.)
\label{fig:res_flux_comp2}}
\end{figure}

\begin{figure}
\plotone{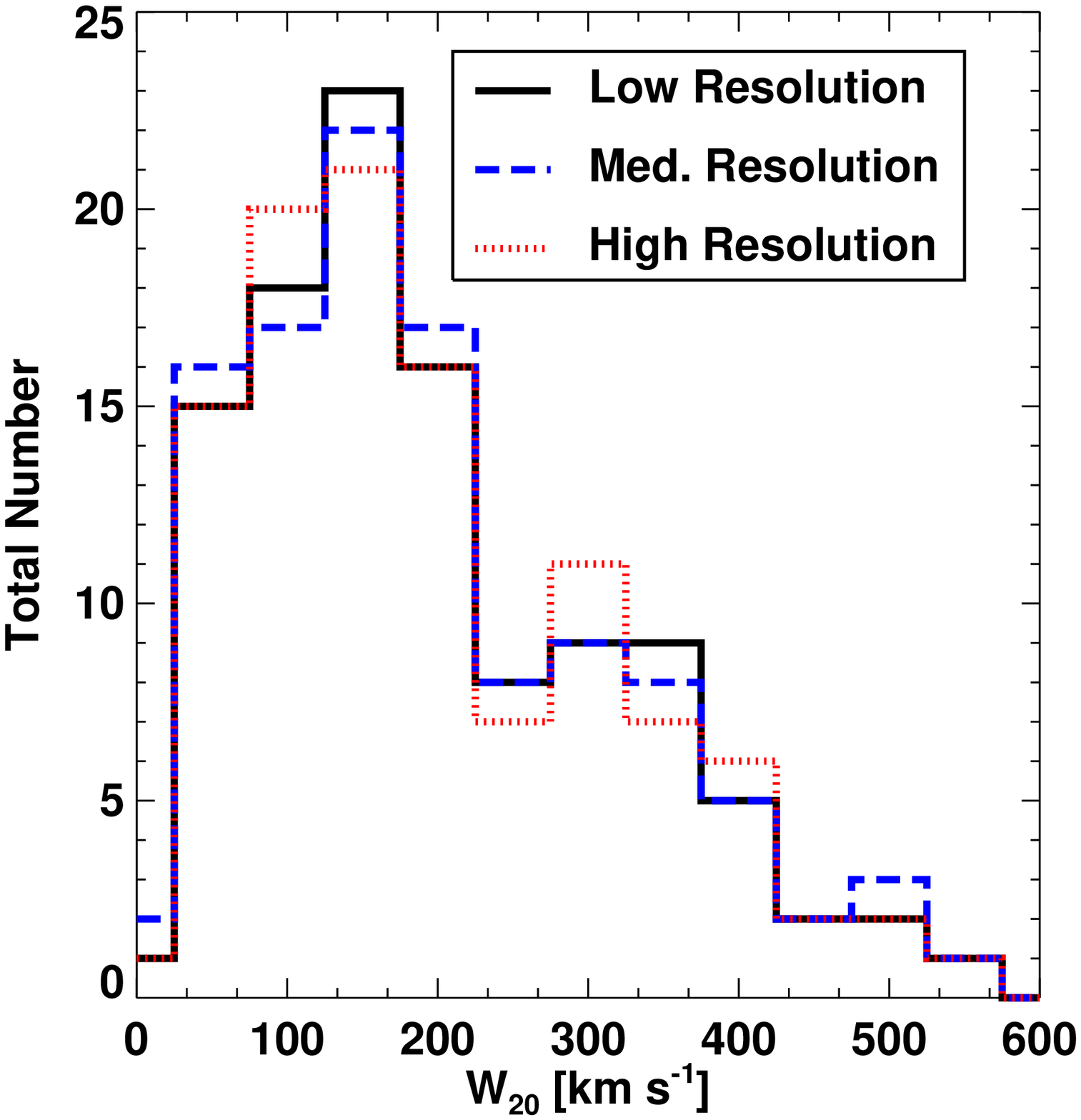}
\caption{A histogram of the W$_{20}$ distribution for
the observed galaxies. The average signal -to-noise for the
three different resolutions is shown. The solid line is the 8.5 \kms\ resolution,
the dashed line is the 1.3 \kms\ resolution, and the dotted line is the 0.65 \kms\
resolution.  \label{fig:res_w20_comp}}
\end{figure}

\begin{figure}
\plotone{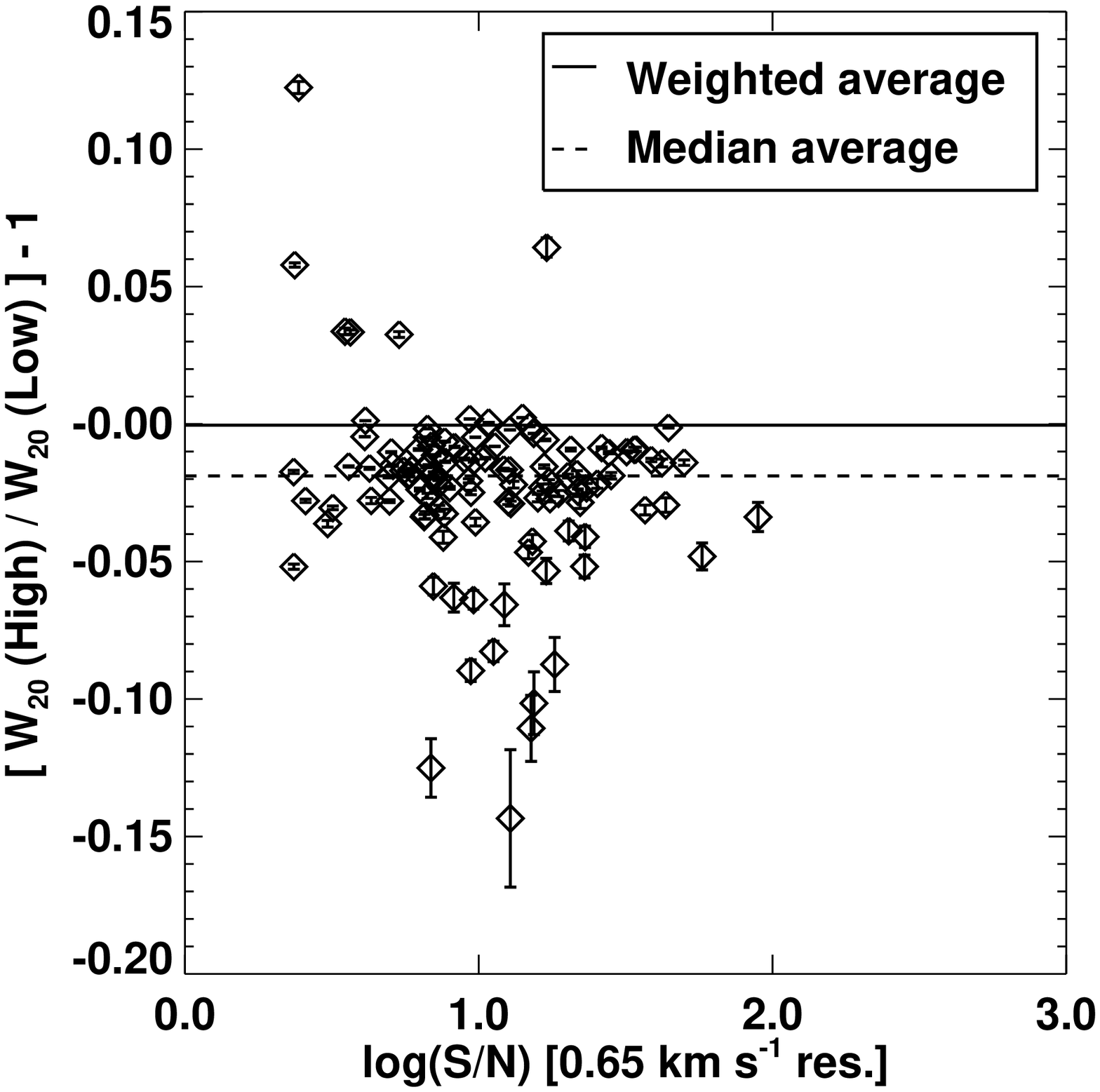}
\caption{
Comparison of the lowest and highest resolution data measured against
the average signal-to-noise of the spectral
line. (Average signal to noise was found through dividing the total flux by the velocity
width of each galaxy.) The solid and dashed red lines simply mark the weighted (by 1/$\sigma^2$)
and unweighted averages of the differences. The weighted average lies at
zero.
\label{fig:res_w20_comp2}}
\end{figure}

\begin{figure}
\plotone{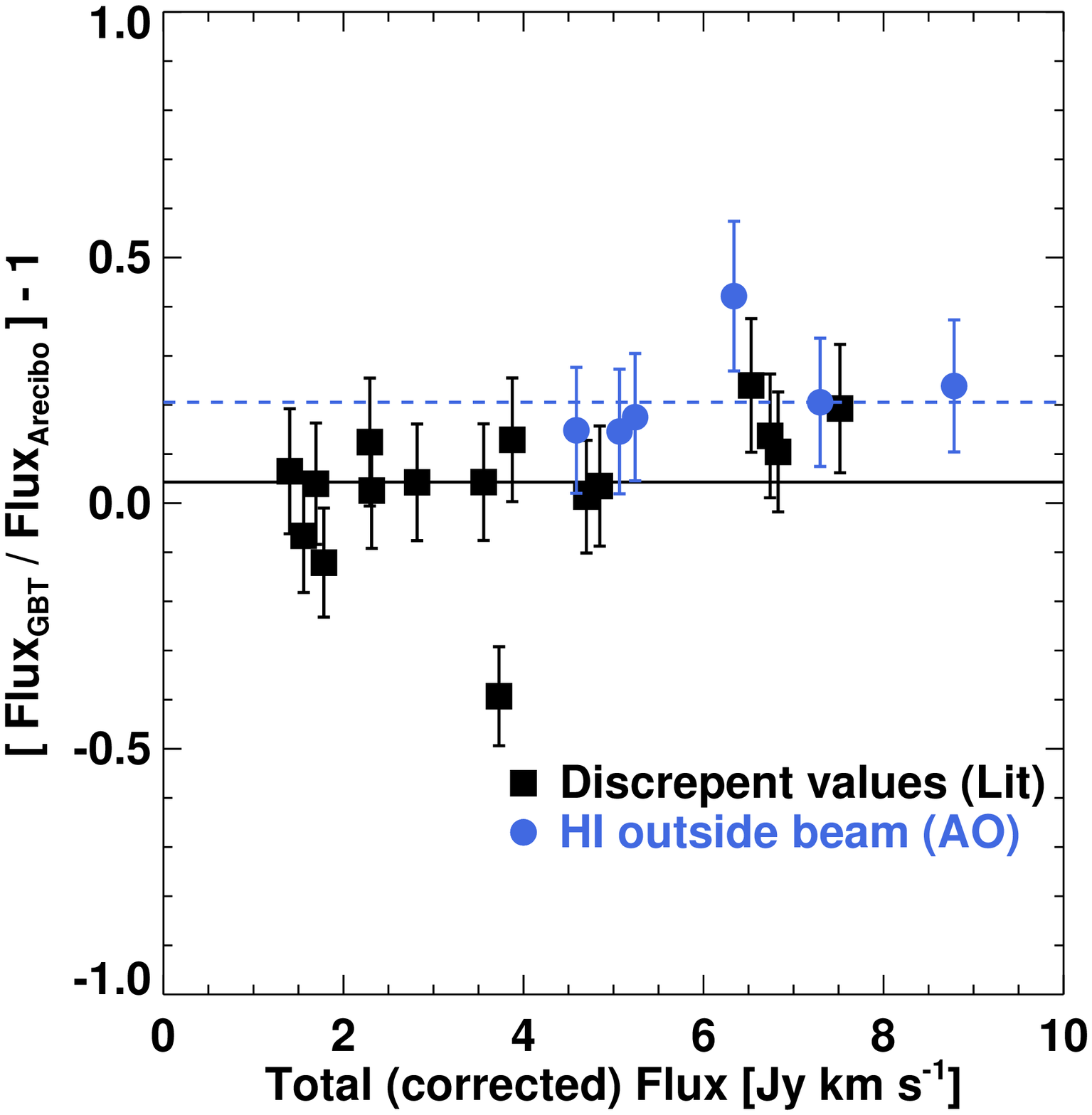}
\caption{Comparison between the fluxes found with the Arecibo and Green Bank Telescopes.
the same is broken into two groups -- objects which we believe our Arecibo measurements
to be correct (squares) and  objects which we believe the flux extends well beyond the Arecibo beam
(circles).  The lines show the median values of the two groups.  See Section 6
for more information on these sources.
\label{fig:gbt} }
\end{figure}

\section{Signal-to-Noise vs. High Resolution}

One interesting observational effect which can be studied with our observations
is the trade-off between high frequency (channel) resolution and high signal-to-noise 
(lower resolution).  
The average S/N is 14, 19, and 46 and the minimum (maximum) S/N seen are 2 (89), 3 (122), and 7 (310)
for the 0.65, 1.3, and 8.5 \kms\ channel$^{-1}$ resolutions, respectively.
The lowest S/N for all three resolutions are from the observations of CAP~0718+06
and the highest from UGC 8091.  Figure~\ref{fig:res_flux_comp2} shows the ratio
of the fluxes found in the high (0.65 \kms) and low (8.5 \kms) resolution plots.

We also compared the flux values obtained for the various
channel resolutions, and found that other than at the very.
low S/N end of the distribution, the overall flux distribution is extremely similar
between the low and high resolution images.  However, looking at the individual
comparisons between the low and high resolution measurements it is (not surprisingly)
clear that the higher the overall signal-to-noise the more consistent the flux measurements
are, with the differences between the measurements being essentially zero once the S/N
reaches $\sim$30.

Figures~\ref{fig:res_w20_comp} and \ref{fig:res_w20_comp2} show a comparison between the
measured (W$_{20}$) velocity widths and the velocity resolution.  The results match
what one would  na\"{i}vely believe --  the higher resolution spectra  provide the most
accurate measurements of the velocity widths.  That is, one can see that the higher
resolution images have, on average, slightly narrower velocity widths than the
lower resolution spectra.

\subsection{Green Bank Telescope Observations}

After the Arecibo observations were analyzed, the fluxes of a number of objects were
found to differ significantly from the literature values of the same sources.  To
check the Arecibo results, we re-observed the most discrepant sources 
(as described in Section~\ref{sec:lit_comp}) using the Robert C. Byrd Green Bank
Telescope (GBT).  All GBT observations took place between 19 April -- 14 May, 2004.
Each galaxy was observed in two separate (linear) polarizations with a 12.5 MHz bandwidth
and 32767 channels/polarization (0.08 \kms\ resolution).  The data was then boxcar
smoothed to 8 \kms\ resolution, to approximate the resolution of the Arecibo observations.
Standard position switching techniques were used for all observations, tracking
the same azimuth and zenith angle for the off-source and on-source observations to
reduce any differences in baseline shapes.
The on- and off-source observations were 5 minutes each with a 10 second on+off
observation of a bright, calibrated, noise diode (the `high-cal') to determine the 
system temperature.

The telescope gain was determined through observing 3-4 standard continuum calibrators 
each night.  As the gain of the GBT at 1.4 GHz does not change significantly until
one observes a source at low zenith angle \citep{ghigo01}, no sources were observed
at zenith angles less than 45\degree.  As a result, the gain found with the
continuum calibrators was applied to all sources without regard to their sky position.  
The variation between the calibration source measurements was between 5--10\%.  The
calibration of the sources should not be trusted to better than this level.

Data reduction took place using the same methods (and routines) as for the Arecibo data,
with the exception that second order baselines were fit to the data when necessary.

The measurements from the GBT observations are given in Table~\ref{tab:gbt}, the 
spectra are shown in Figure~\ref{fig:HI_profs}.  A comparison between the GBT and Arecibo
measurements is shown in Figure~\ref{fig:gbt}.  The average offset between the GBT
and Arecibo measurements of sources which do not have \HI\ extended outside
the 3.5$^\prime$ Arecibo beam (the black symbols in Figure~\ref{fig:gbt})
is 4$\pm$20\%, essentially zero.

\begin{figure}
\plotone{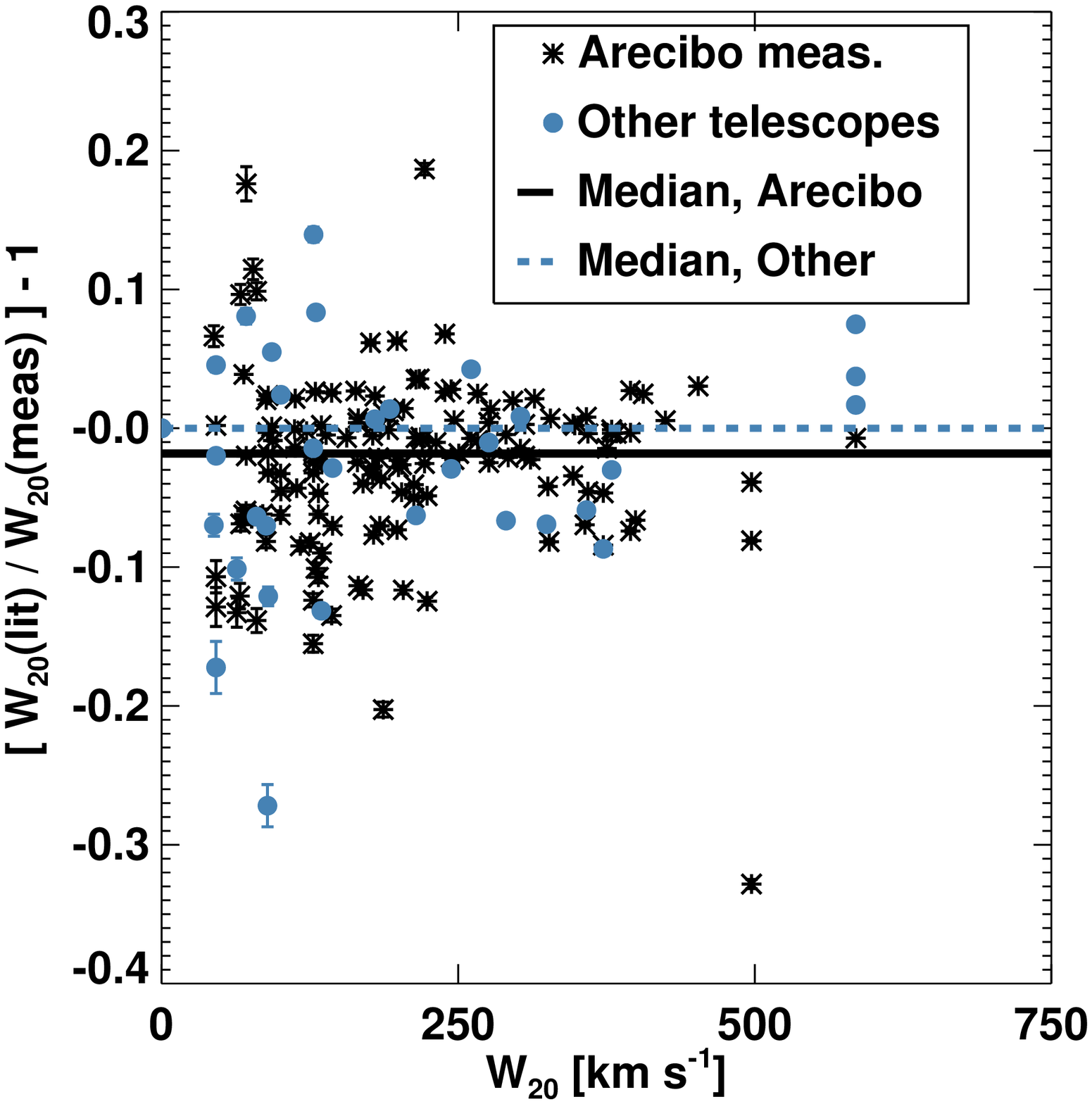}
\caption{Comparison between the velocity widths of our survey
with the values found in the literature. To insure accuracy in the comparisons,
the data from our survey was smoothed to the same resolution as that commonly found
in previous observations (8.5 \kms\ at 1420 MHz). Additionally, the literature sample
was broken into two categories - previous Arecibo Observations (all done before the
upgrade) and those observations taken at other radio telescopes.
No significant different from zero is seen for the
mean of the difference between out measurements and those in the literature.
The comparison between the value of UGC 9007 found by our survey and that found in \citep{1985ApJS...57..423}
is not shown on the plot.  It should lie at W$_{20}$=76 \kms, difference=1.35.
\label{fig:obs_lit_w20_comp}}
\end{figure}

\section{Comparison with Literature and GBT Values\label{sec:lit_comp}}

\begin{figure}
\plotone{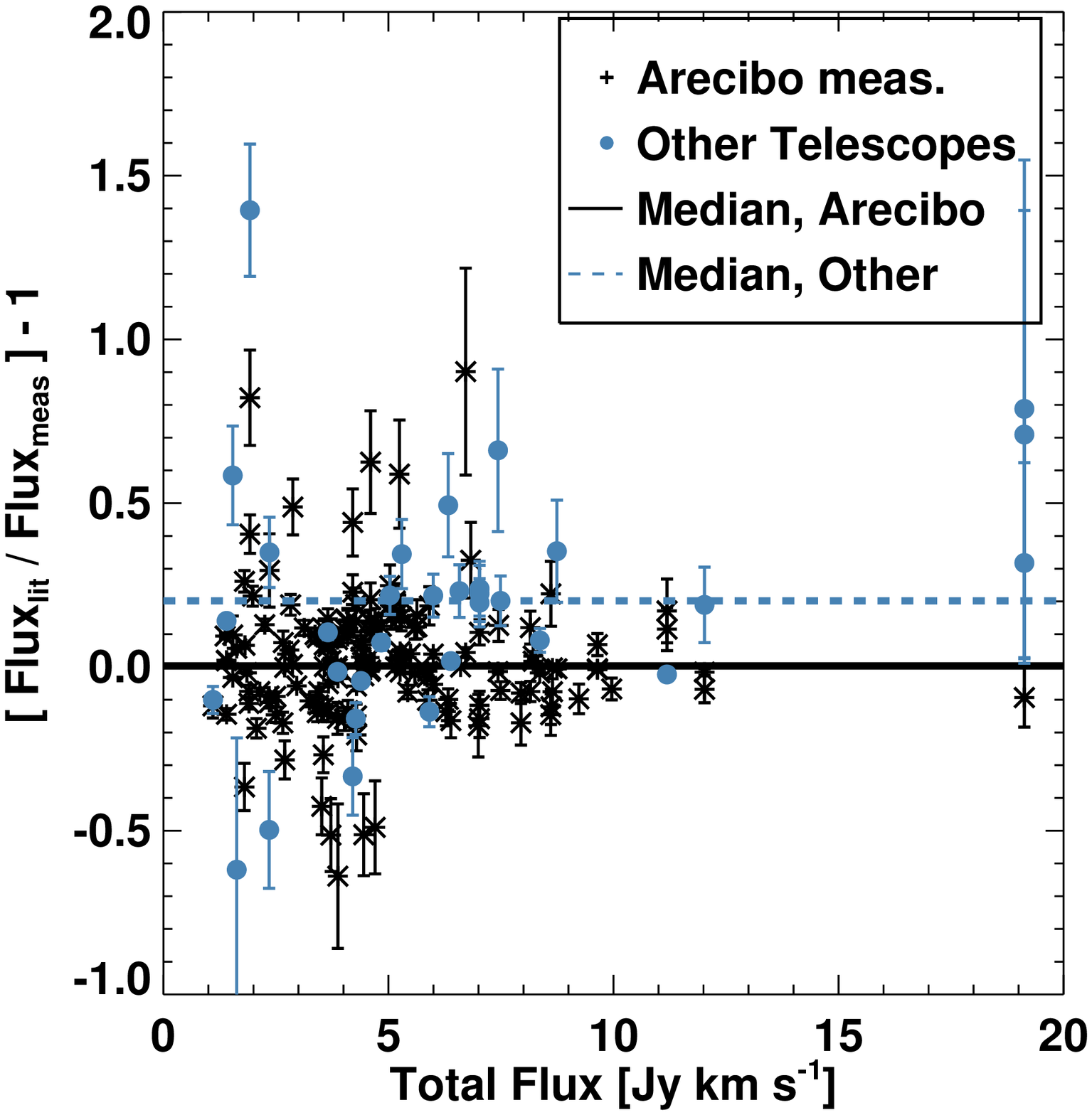}
\caption{A comparison between the fluxes found with our survey
and the values found in the literature. To insure accuracy in the comparison,
the data from our survey was smoothed to the same resolution as that commonly found
in previous observations (8.5 \kms\ at 1420 MHz). Additionally, the literature sample
was broken into two categories - previous Arecibo Observations (all done before the
upgrade) and those observations taken at other radio telescopes.  This plot
shows a a direct comparison between our data and that in the literature.
The mean of the unweighted flux differences
lie at 0.00 and 0.19 for the Arecibo and non-Arecibo samples, respectively.
\label{fig:obs_lit_comp}}
\end{figure}

\begin{figure}
\plotone{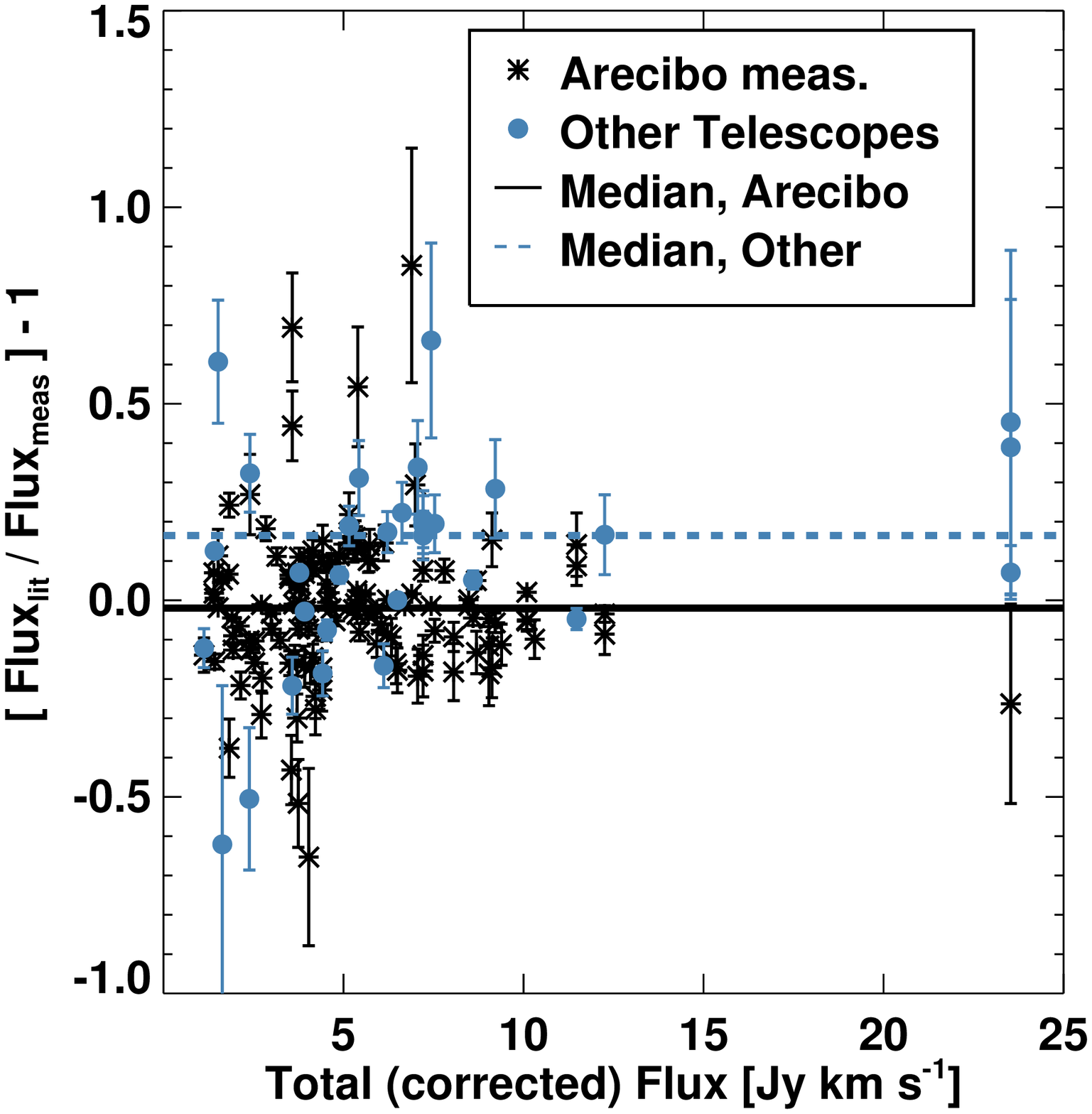}
\caption{Comparison between the fluxes found with our survey
and the values found in the literature. To insure accuracy in the comparisons,
the data from our survey was smoothed to the same resolution as that commonly found
in previous observations (8.5 \kms\ at 1420 MHz). Additionally, the literature sample
was broken into two categories - previous Arecibo Observations (all done before the
upgrade) and those observations taken at other radio telescopes.  This plot
compares our data after it has been corrected for the beam width
(see Section~\ref{sec:lit_comp}).  
The unweighted medians lie at $-$0.02 and
0.14 for the Arecibo and non-Arecibo samples, respectively.
\label{fig:obs_lit_comp2}}
\end{figure}

 Figure~\ref{fig:obs_lit_w20_comp} shows a
comparison between the velocity width (W$_{20}$) in our survey versus all values found in the literature
(searched through NED -- the NASA Extragalactic Database, and listed in
Table~\ref{tab:lit}).  Similar
plots are shown in Figures~\ref{fig:obs_lit_comp} and \ref{fig:obs_lit_comp2}
for the flux measurements, using the final (gain, beam, and pointing-corrected)
values in the literature for the comparison.  In all plots the literature
values are split into two categories -- measurements previously
made at Arecibo, and measurements made at any other telescope.
Errors for the literature value typically came directly from the
published value.  In the cases where no value was published, the
flux error was estimated from multiplying the reported velocity
width by the reported r.m.s. error.  When no error was reported
for the measured velocity widths, the error was estimated to be equal to
half a velocity channel.  Errors used are given in Table~\ref{tab:lit},
and a description of the comparisons is below.

\subsection{Velocity Width Comparisons}

For the velocity width measurement comparisons (Figure~\ref{fig:obs_lit_w20_comp}),
the weighted average and
mean of the differences between our observations and those
found in the literature are zero regardless of the telescope
used for the observations.  This shows there are no systematic
errors affecting our results.  In fact, there are only 3
observations (out of a total of 186) which vary by more than
20\% from the value our measured values.  In each case follow-up
observations taken with the GBT confirmed our Arecibo measurements.
The `errant' literature values are:

\begin{itemize}
\item {\bf UGC 3118:} There are two measurements of this galaxy
in the literature, one of which agrees with our Arecibo and GBT measurements
to within the errors, and one which does not.  The `errant'
measurement was made by \citet{1992ApJS...81....5} with the
Green Bank 300$^\prime$ telescope.  Schneider, et.al list
their observation as potentially confused with another source,
which likely explains the difference in measurements.

\item{\bf UGC 9007:} Three values for the \HI\ properties of
UGC 9007 can be found in the literature, two which match that found
by our Arecibo and GBT observations and one which does not.  The value reported by
\citet{1985ApJS...57..423} is roughly 100 \kms\ larger than that found by all
other observations.  This, combined with the fact that the
Bothun et al. W$_{50}$ measurement is considerable smaller than
their W$_{20}$ measurement and is consistent with the other values found
in the literature makes it likely that their high value for W$_{20}$
is in error.

\item{\bf UGC 11017:} There are two published values for velocity width
of UGC 11017, only one of which does not match that found in
our survey.  This measurement, which is lower by 40 \kms\ from the
the other available measurements, was taken by \citet{1991AAS...87..425} using the
Arecibo telescope.  Examining their profile shows the observation to
be extremely low signal-to-noise, making their measured values suspect.
\end{itemize}

\subsection{Total Flux Comparisons}
\label{sec:flux}

For the flux measurements, the weighted average
of the differences between our observations and those
found in the literature (using the gain, beam, and pointing-corrected 
values in the literature) are zero regardless of the telescope
used for the observations when comparing direct observations.
However, the median difference between measurements in
the literature and that found by us for non-Arecibo measurements
differ by 20\%, with the literature values being
higher on average than our measurements (Figures~\ref{fig:obs_lit_comp}
and \ref{fig:obs_lit_comp2}).

To determine if the offset between the non-Arecibo
measurements is due to many of the
studies galaxies having \HI\ emission extending beyond
the Arecibo beam, we applied a beam dilution correction
to all of the galaxies on our study.  The beam dilution correction
is based on Shostak's 1978 equation for the
fraction of the total flux integral measured by a beam centered on a galaxy:
\[ f\;=\;{{\sum_{j=1,2} \left[ {{a_j \theta_j^2}\over{\left({1+\theta_j^2/\theta_B^2}\right)^{1/2}
\;\left({1+\theta_j^2/\theta_B^2\;cos^2i}\right)^{1/2}}}\right]} \over
{\sum_{j=1,2} a_j \theta_j^2}}.\]
Here, $\theta_1 \sim D_{70}/1.1$ with $D_{70} = 1.2a$ (arcmin),
$a_2 = -0.6a_1$ and $\theta_2 =0.23\theta_1$ \citep{hewitt83}.
Additionally, the inclination is derived using the method described in
\citet{1984AJ.....89..758} assuming galaxies are oblate spheroids:
\[ cos^2i\;=\;\left({r_i^2\;-\;p^2}\right)\left({1\;-\;p^2}\right)^{-1}\]
where $r_i = b/a$, $p = c/a$, $a$, $b$, and $c$ are the
spheroid's three axiis, and $p$ is determined from the galaxy type.
The result of applying this correction to the data is shown in
Figure~\ref{fig:obs_lit_comp2}.  Again the weighted difference between our observations
with the corrected flux and those found in the literature are zero
regardless of the telescope used for the observations.
In this case, though, the median difference between measurements in
the literature and that found by us for non-Arecibo measurements
is lowered to 16\%, with the literature values again being
higher than our measurements (Figure~\ref{fig:obs_lit_comp2}).
Consequently, while it seems likely that extended \HI\ for
some of the observed galaxies is responsible for the high differences
in the median flux values between our measurements and those
taken at telescopes with larger beams, the flux correction did
not adequately (and possibly not accurately) correct for the extended
\HI.

To more thoroughly examine the flux differences found, we examined
all galaxies with any published flux measurement differing from
our measured value by at least $\pm$20\%.  Of the 26 galaxies
which fall into this category, 25 were observed with the GBT.
The last galaxy, UGC 12695, had VLA synthesis measurements which agree with
our Arecibo observations, and no GBT follow-up was warranted.
A complete listing of the 26 galaxies follows:

\begin{itemize}
\item 13 galaxies (UGC~0941, UGC~2885, UGC~3118, UGC~5160, UGC~5215, 
UGC~5218, UGC~6886, UGC~8503, UGC~9803, UGC~9450, UGC~11017, UGC~11627, UGC~12094) have
values in the literature which vary significantly from our Arecibo measurements.
In each case, follow-up observations with the GBT confirmed the Arecibo
values.
\item 7 galaxies (UGC~8091, UGC~9965, UGC~10351, UGC~10628, UGC~11461, 
UGC~11578, UGC~11992) have all flux values measured with a telescope with a larger 
beam, including our GBT measurements, to be larger than the values
obtained at Arecibo (both by ourself and in the literature.
As none of these galaxies have another
source within 20$^\prime$/1000 \kms, it is safe to assume
that these objects have \HI\ gas which extends well beyond the
Arecibo beam, and that while our observations should be repeatable
with the Arecibo telescope, the measured fluxes do not accurately
reflect the total \HI\ mass of the objects.
\item Four other galaxies (UGC~2975, UGC~3476, UGC~4660, UGC~6464)
with flux values in the literature discrepant from our Arecibo measurements
were also observed with the GBT.  Unfortunately, the follow-up observations
did not reach a sufficient signal-to-noise to accurately determine the galaxies'
fluxes with the GBT. 

The only literature values for UGC~2975 and UGC~6464 
are from the pre-upgrade Arecibo telescope.  The values found for UGC~6464 
straddle those found with our observations, while the one value previously published
for UGC~2975 is smaller than that found with our observations.  In light of the
extremely accurate fluxes found for all other sources in this survey, we are confident
that our recent measurements of these objects are correct.

UGC 4660 and UGC 3476 have literature values from the 300' Green Bank telescope
which are larger than any Arecibo measurements taken of the galaxies.  As a result, 
it is likely that the \HI\ gas of these objects extends outside the Arecibo beam.
\item UGC 12695 has 3 single dish observations and one interferometric (VLA) observation
in the literature.  The two pre-upgrade Arecibo measurements are higher than that found
with our observations by 23--44\%, while the Nan\c{c}ay measurement is 33\% lower than
ours.  However, the published VLA value agrees with ours to within a few percent,
again giving confidence that our measurements are accurate.
\end{itemize}

Compiling the above results shows that of the 108 galaxies observed for this
project, only 9 have \HI\ gas which extends beyond the Arecibo beam.
As a result, measurements for these objects, while repeatable with the Arecibo
305m telescope, are not accurate representations of the total amount of
neutral Hydrogen within the galaxies.  Each of these objects is clearly marked
in Table~\ref{tab:gals}.

\section{Conclusion}

We have created a very reliable reference catalog of \HI\ spectral
line observations.  Of the 108 galaxies observed, only 9 have measured fluxes indicating
significant \HI\ gas extending beyond the Arecibo beam.  As a result, our
catalog currently contains the most accurate \HI\ line measurements and
spectra available for the remaining 99 galaxies.  This catalog should be
an extremely useful, well-defined reference catalog for anyone interested in
21-cm spectral line work.

All data and results from this survey are available online at both
\url{http://www.naic.edu/$\sim$astro/HIsurvey} and
\url{http://www.gb.nrao.edu/$\sim$koneil/HIsurvey}.

\onecolumn
\begin{deluxetable}{lcccclccclccc}
\rotate
\tabletypesize{\scriptsize}
\tablewidth{0pt}
\tablecaption{Final Results, Broken Down by Resolution \label{tab:gals}}
\tablehead{
& & & \multicolumn{3}{c}{\bf 0.65 km s$^{-1}$ Resolution} &
\multicolumn{3}{c}{\bf 1.3 km s$^{-1}$ Resolution} &
\multicolumn{3}{c}{\bf 8.5 km s$^{-1}$ Resolution} \\
\colhead{Name} & \colhead{RA} & \colhead{Dec} & \colhead{V$_{HEL}$} &
\colhead{Flux} & \colhead{W$_{20}$} & \colhead{W$_{50}$} &
\colhead{Flux} & \colhead{W$_{20}$} & \colhead{W$_{50}$} &
\colhead{Flux} & \colhead{W$_{20}$} & \colhead{W$_{50}$} \\
& \multicolumn{2}{c}{J2000}   & \colhead{[\kms]} & 
\colhead{[Jy \kms]} & \colhead{[\kms]} & \colhead{[\kms]} &
\colhead{[Jy \kms]} & \colhead{[\kms]} & \colhead{[\kms]} &
\colhead{[Jy \kms]} & \colhead{[\kms]} & \colhead{[\kms]} 
}
\startdata
     UGC   199 & 00:20:51.7 & 12:51:22.0 &     1800 (3) &     3.6 (0.2) &    117 (7) &    111 (3) &     3.7 (0.2) &    122 (4) &    112 (3) &     3.6 (0.2) &    122 (4) &    112 (4) \\
      NGC 0257 & 00:48:01.6 & 08:17:48.0 &     5274 (2) &     9.9 (0.5) &    408 (11) &    404 (1) &     9.9 (0.5) &    422 (10) &    404 (3) &     9.9 (0.5) &    415 (4) &    406 (4) \\
     UGC   317 & 00:31:43.2 & 00:54:03.0 &     5361 (1) &     1.1 (0.1) &     37 (6) &     33 (4) &     1.1 (0.1) &     44 (2) &     35 (2) &     1.1 (0.1) &     44 (5) &     38 (4) \\
     UGC   628 & 01:00:51.9 & 19:28:33.0 &     5446 (5) &     4.4 (0.2) &    241 (4) &    232 (5) &     4.4 (0.2) &    243 (7) &    234 (3) &     4.4 (0.2) &    244 (4) &    234 (4) \\
     UGC   941 & 01:23:32.4 & 06:57:38.0 &     2729 (4) &     3.6 (0.2) &     87 (8) &     85 (7) &     3.6 (0.2) &    104 (4) &     80 (3) &     3.6 (0.2) &     89 (4) &     82 (4) \\
     UGC   989 & 01:25:29.6 & 07:33:41.0 &     2785 (2) &     5.4 (0.3) &    161 (11) &    153 (4) &     5.4 (0.3) &    164 (5) &    154 (3) &     5.4 (0.3) &    165 (4) &    154 (4) \\
     UGC  1085 & 01:31:18.8 & 07:47:16.0 &      651 (3) &     2.1 (0.1) &     60 (7) &     55 (2) &     2.1 (0.1) &     74 (5) &     56 (6) &     2.1 (0.1) &     65 (4) &     60 (4) \\
     UGC  1246 & 01:47:00.7 & 12:24:21.0 &      803 (1) &     8.6 (0.4) &    113 (3) &     99 (2) &     8.7 (0.4) &    117 (2) &    100 (1) &     8.6 (0.4) &    113 (4) &    103 (4) \\
     UGC  1519 & 02:02:16.8 & 19:10:48.0 &     2346 (3) &     2.5 (0.1) &    129 (3) &    121 (4) &     2.4 (0.1) &    135 (2) &    122 (4) &     2.5 (0.1) &    135 (4) &    125 (4) \\
     UGC  2008 & 02:32:32.2 & 31:36:35.0 &     5043 (4) &     2.7 (0.1) &    139 (8) &    128 (8) &     2.7 (0.1) &    150 (9) &    130 (4) &     2.7 (0.1) &    143 (4) &    133 (4) \\
     UGC  2339 & 02:51:23.6 & 16:02:30.0 &    16467 (11) &     4.5 (0.2) &    497 (3) &    469 (3) &     4.5 (0.2) &    500 (16) &    472 (17) &     4.5 (0.2) &    497 (7) &    475 (7) \\
     UGC  2432 & 02:57:26.7 & 10:08:12.0 &      758 (1) &     5.9 (0.3) &    114 (1) &    105 (0) &     5.9 (0.3) &    115 (1) &    104 (1) &     5.9 (0.3) &    116 (4) &    105 (4) \\
        AC HVC &     -753 (1) &  ...     &  ...  &  ...  &     0.7 (0.1) &     42 (1) &     36 (2) &   ...    &  ...  &  ...  \\
     UGC  2602 & 03:14:10.1 & 16:29:08.0 &    10092 (8) &     6.0 (0.3) &    302 (17) &    290 (11) &     5.9 (0.3) &    307 (16) &    290 (4) &     5.9 (0.3) &    302 (7) &    290 (4) \\
     UGC  2809 & 03:39:33.2 & 19:47:03.0 &     1287 (1) &     9.9 (0.5) &    154 (2) &    144 (1) &     9.9 (0.5) &    156 (2) &    144 (1) &     9.9 (0.5) &    156 (4) &    146 (4) \\
     UGC  2885 & 03:53:02.4 & 35:35:22.0 &     5803 (7) &    19.2 (1.0) &    580 (4) &    567 (8) &    19.0 (1.0) &    591 (7) &    567 (3) &    19.2 (1.0) &    585 (4) &    572 (4) \\
     UGC  2905 & 03:57:00.1 & 16:31:21.0 &      295 (1) &     4.1 (0.2) &     75 (7) &     65 (3) &     4.0 (0.2) &     82 (1) &     68 (1) &     4.1 (0.2) &     75 (4) &     68 (4) \\
     UGC  2922 & 04:00:46.0 & 17:35:00.0 &     5220 (2) &     4.6 (0.2) &    369 (4) &    364 (7) &     4.7 (0.2) &    380 (7) &    361 (3) &     4.6 (0.2) &    375 (5) &    364 (5) \\
     UGC  2975 & 04:11:00.4 & 26:36:46.0 &    19945 (17) &     2.0 (0.1) &    316 (10) &    282 (2) &     2.0 (0.1) &    327 (3) &    284 (2) &     2.0 (0.1) &    325 (11) &    285 (16) \\
     UGC  3044 & 04:26:50.0 & 29:56:52.0 &     5231 (7) &     2.7 (0.1) &    248 (8) &    240 (10) &     2.6 (0.1) &    254 (6) &    240 (6) &     2.7 (0.1) &    250 (5) &    240 (4) \\
     UGC  3118 & 04:38:59.0 & 05:37:11.0 &     8314 (5) &     1.6 (0.1) &     78 (4) &     74 (1) &     1.6 (0.1) &     99 (7) &     75 (1) &     1.6 (0.1) &     89 (11) &     75 (4) \\
     UGC  3169 & 04:47:45.0 & 09:35:43.0 &     8800 (8) &     3.0 (0.2) &    378 (0) &    382 (0) &     3.0 (0.2) &    400 (1) &    370 (1) &     2.9 (0.1) &    399 (4) &    374 (4) \\
     UGC  3247 & 05:06:38.0 & 08:40:27.0 &     3365 (7) &     5.2 (0.3) &    220 (0) &    208 (0) &     5.1 (0.3) &    239 (1) &    177 (1) &     5.2 (0.3) &    223 (4) &    209 (4) \\
     UGC  3296 & 05:21:21.8 & 04:53:16.0 &     4267 (3) &     8.6 (0.4) &    342 (0) &    332 (0) &     8.8 (0.4) &    361 (1) &    335 (1) &     8.6 (0.4) &    346 (4) &    333 (4) \\
     UGC  3330 & 05:36:46.9 & 14:25:21.0 &     5238 (3) &     4.6 (0.2) &    211 (11) &    189 (18) &     4.5 (0.2) &    216 (7) &    185 (11) &     4.6 (0.2) &    212 (7) &    188 (4) \\
 UGC  3476\dag & 06:30:29.2 & 33:18:07.0 &      472 (5) &     7.4 (0.4) &    128 (13) &    104 (8) &     7.3 (0.4) &    150 (3) &    106 (5) &     7.4 (0.4) &    130 (4) &    119 (4) \\
     UGC  3564 & 06:50:36.2 & 16:21:20.0 &     2546 (0) &     5.6 (0.3) &     97 (0) &     86 (1) &     5.6 (0.3) &    103 (1) &     90 (1) &     5.6 (0.3) &    100 (4) &     91 (4) \\
     UGC  3582 & 06:53:00.3 & 12:11:17.0 &     8392 (10) &     2.8 (0.1) &    177 (28) &    165 (23) &     2.8 (0.1) &    189 (18) &    164 (2) &     2.8 (0.1) &    183 (11) &    166 (4) \\
     UGC  3621 & 06:59:04.4 & 14:17:42.0 &     2335 (2) &     9.6 (0.5) &    176 (1) &    165 (2) &     9.6 (0.5) &    177 (3) &    164 (1) &     9.6 (0.5) &    177 (4) &    166 (4) \\
     UGC  3755 & 07:13:51.5 & 10:31:19.0 &      318 (4) &     8.2 (0.4) &     67 (7) &     77 (6) &     8.1 (0.4) &    111 (3) &     79 (5) &     8.2 (0.4) &     71 (4) &     80 (4) \\
     UGC  3767 & 07:14:47.7 & 06:46:50.0 &     5807 (4) &     4.0 (0.2) &    261 (4) &    254 (3) &     4.0 (0.2) &    264 (10) &    254 (6) &     4.0 (0.2) &    266 (4) &    257 (4) \\
   CAP 0718+06 & 07:21:00.2 & 06:18:19.0 &    11140 (6) &     1.4 (0.1) &    207 (10) &    207 (11) &     1.4 (0.1) &    209 (11) &    192 (11) &     1.4 (0.1) &    211 (16) &    195 (8) \\
     UGC  3946 & 07:37:59.7 & 03:18:58.0 &     1198 (6) &    12.0 (0.6) &    123 (13) &    112 (4) &    12.1 (0.6) &    143 (9) &    114 (1) &    12.0 (0.6) &    127 (4) &    113 (4) \\
     UGC  4117 & 07:57:26.0 & 35:56:21.0 &      773 (1) &     4.1 (0.2) &     80 (5) &     73 (2) &     4.1 (0.2) &     88 (1) &     76 (1) &     4.1 (0.2) &     85 (4) &     76 (4) \\
     UGC  4131 & 07:59:11.6 & 31:48:28.0 &    17728 (24) &     3.4 (0.2) &    464 (14) &    448 (19) &     3.4 (0.2) &    467 (32) &    451 (25) &     3.4 (0.2) &    466 (16) &    448 (25) \\
     UGC  4180 & 08:02:36.5 & 27:26:15.0 &     5229 (2) &     5.3 (0.3) &    354 (2) &    347 (9) &     5.2 (0.3) &    356 (4) &    348 (5) &     5.3 (0.3) &    359 (4) &    350 (4) \\
     UGC  4385 & 08:23:52.0 & 14:45:07.0 &     1967 (5) &     6.6 (0.3) &    185 (7) &    168 (2) &     6.6 (0.3) &    190 (9) &    169 (5) &     6.6 (0.3) &    188 (4) &    170 (4) \\
 LSBC F704-V01 & 08:24:51.7 & 09:13:29.0 &     6018 (5) &     2.0 (0.1) &    135 (10) &    130 (13) &     2.0 (0.1) &    141 (5) &    131 (5) &     2.0 (0.1) &    139 (4) &    132 (4) \\
 UGC  4660\dag & 08:54:24.2 & 34:33:20.0 &     2202 (2) &     6.0 (0.3) &     77 (4) &     69 (2) &     6.0 (0.3) &     80 (2) &     69 (1) &     6.0 (0.3) &     80 (4) &     73 (4) \\
     UGC  4780 & 09:06:39.4 & 19:20:10.0 &     3287 (6) &     8.1 (0.4) &    164 (15) &    139 (5) &     8.2 (0.4) &    169 (2) &    140 (1) &     8.1 (0.4) &    165 (4) &    142 (4) \\
     UGC  4955 & 09:20:19.3 & 25:16:25.0 &     6463 (4) &     2.1 (0.1) &    209 (0) &    202 (0) &     2.1 (0.1) &    214 (1) &    206 (1) &     2.1 (0.1) &    217 (4) &    207 (4) \\
     UGC  5030 & 09:26:36.2 & 07:57:15.0 &     2153 (2) &     3.4 (0.2) &    300 (3) &    295 (7) &     3.3 (0.2) &    305 (12) &    295 (11) &     3.4 (0.2) &    306 (8) &    297 (4) \\
     UGC  5160 & 09:40:41.7 & 11:53:18.0 &     6660 (13) &     2.9 (0.1) &    399 (21) &    398 (25) &     2.8 (0.1) &    402 (25) &    390 (24) &     2.9 (0.1) &    405 (19) &    397 (8) \\
     UGC  5215 & 09:45:14.3 & 09:06:35.0 &     5485 (1) &     6.8 (0.3) &    420 (0) &    418 (0) &     6.8 (0.3) &    425 (1) &    416 (1) &     6.8 (0.3) &    424 (4) &    417 (4) \\
     UGC  5218 & 09:45:30.2 & 06:22:37.0 &     3088 (1) &     6.7 (0.3) &    195 (4) &    185 (2) &     6.7 (0.3) &    200 (1) &    186 (1) &     6.7 (0.3) &    199 (4) &    188 (4) \\
     UGC  5326 & 09:55:24.6 & 33:15:45.0 &     1412 (8) &     4.3 (0.2) &    131 (6) &    112 (6) &     4.3 (0.2) &    128 (8) &    111 (4) &     4.3 (0.2) &    132 (11) &    115 (4) \\
     UGC  5358 & 09:58:47.2 & 11:23:19.0 &     2914 (1) &     4.4 (0.2) &    216 (0) &    203 (0) &     4.5 (0.2) &    222 (1) &    204 (1) &     4.4 (0.2) &    218 (4) &    207 (4) \\
     UGC  5440 & 10:05:35.9 & 04:16:45.0 &    18958 (14) &     3.0 (0.2) &    537 (34) &    537 (35) &     3.1 (0.2) &    513 (19) &    484 (20) &     2.9 (0.1) &    508 (13) &    498 (13) \\
     UGC  5629 & 10:24:12.8 & 21:03:01.0 &     1238 (1) &     2.9 (0.1) &    126 (4) &    116 (4) &     2.8 (0.1) &    129 (3) &    116 (4) &     2.9 (0.1) &    129 (4) &    118 (4) \\
     UGC  5651 & 10:26:25.6 & 17:30:38.0 &     5567 (2) &     5.3 (0.3) &    224 (10) &    218 (1) &     5.2 (0.3) &    234 (4) &    219 (1) &     5.2 (0.3) &    231 (4) &    221 (4) \\
     UGC  5852 & 10:44:12.1 & 06:45:33.0 &     6174 (4) &    11.2 (0.6) &    380 (23) &    362 (4) &    11.2 (0.6) &    385 (8) &    363 (4) &    11.2 (0.6) &    379 (7) &    364 (4) \\
     UGC  6018 & 10:54:05.8 & 20:38:41.0 &     1294 (2) &     1.9 (0.1) &     61 (11) &     57 (6) &     1.9 (0.1) &     70 (17) &     57 (3) &     1.9 (0.1) &     69 (5) &     61 (4) \\
     UGC  6421 & 11:24:25.7 & 27:27:23.0 &     1502 (1) &     7.9 (0.4) &    175 (2) &    169 (0) &     8.0 (0.4) &    181 (5) &    170 (1) &     7.9 (0.4) &    178 (4) &    171 (4) \\
     UGC  6464 & 11:28:09.3 & 16:55:14.0 &     1075 (5) &     3.5 (0.2) &    154 (1) &    141 (8) &     3.4 (0.2) &    168 (5) &    144 (10) &     3.6 (0.2) &    169 (8) &    148 (4) \\
     UGC  6476 & 11:28:36.0 & 23:24:15.0 &     7328 (9) &     3.7 (0.2) &    321 (4) &    319 (1) &     3.6 (0.2) &    328 (1) &    319 (5) &     3.7 (0.2) &    326 (4) &    319 (5) \\
   \obc\ U01-4 & 11:38:25.7 & 17:05:03.0 &     3452 (3) &     1.8 (0.1) &    197 (0) &    177 (3) &     1.8 (0.1) &    192 (7) &    176 (10) &     1.8 (0.1) &    191 (4) &    182 (4) \\
     UGC  6872 & 11:53:50.9 & 33:21:55.0 &     3203 (5) &     5.7 (0.3) &    216 (6) &    198 (30) &     5.7 (0.3) &    220 (11) &    198 (11) &     5.6 (0.3) &    221 (29) &    201 (7) \\
     UGC  6886 & 11:55:11.3 & 06:10:09.0 &     6973 (20) &     4.6 (0.2) &    384 (11) &    383 (22) &     4.6 (0.2) &    404 (50) &    377 (1) &     4.6 (0.2) &    395 (7) &    383 (7) \\
     UGC  7302 & 12:16:49.4 & 30:16:13.0 &     3838 (2) &     3.9 (0.2) &     91 (6) &     82 (1) &     3.9 (0.2) &     95 (5) &     83 (1) &     3.9 (0.2) &     93 (4) &     84 (4) \\
     UGC  7666 & 12:31:37.7 & 14:51:38.0 &     2556 (10) &     2.2 (0.1) &    129 (18) &    111 (2) &     2.3 (0.1) &    151 (14) &    116 (7) &     2.2 (0.1) &    138 (8) &    117 (4) \\
       Malin 1 & 12:36:59.2 & 14:19:50.0 &    24768 (17) &     2.1 (0.1) &    417 (16) &    396 (14) &     2.1 (0.1) &    514 (10) &    331 (8) &     2.1 (0.1) &    372 (7) &    332 (5) \\
     UGC  7976 & 12:49:15.8 & 04:39:27.0 &     2666 (1) &     6.4 (0.3) &     92 (11) &     80 (1) &     6.4 (0.3) &     94 (1) &     81 (1) &     6.4 (0.3) &     92 (4) &     82 (4) \\
UGC  8091\ddag & 12:58:40.0 & 14:13:00.0 &      213 (0) &     7.0 (0.4) &     44 (2) &     44 (1) &     7.0 (0.4) &     55 (2) &     44 (2) &     7.0 (0.4) &     45 (4) &     45 (4) \\
     UGC  8249 & 13:10:26.5 & 24:55:15.0 &     2541 (1) &     7.5 (0.4) &    187 (14) &    178 (2) &     7.4 (0.4) &    189 (5) &    179 (2) &     7.5 (0.4) &    192 (4) &    181 (4) \\
     UGC  8503 & 13:30:44.9 & 32:45:38.0 &     4676 (4) &     3.7 (0.2) &     72 (6) &     65 (2) &     3.7 (0.2) &     83 (8) &     69 (2) &     3.7 (0.2) &     75 (5) &     71 (4) \\
     UGC  8516 & 13:31:52.5 & 20:00:01.0 &     1020 (3) &     3.6 (0.2) &    125 (11) &    114 (2) &     3.7 (0.2) &    126 (12) &    115 (2) &     3.7 (0.2) &    128 (4) &    115 (4) \\
     UGC  8896 & 13:58:38.5 & 07:12:58.0 &     4401 (9) &     5.1 (0.3) &    303 (18) &    291 (20) &     5.1 (0.3) &    312 (21) &    298 (16) &     5.1 (0.3) &    311 (4) &    298 (4) \\
     UGC  8904 & 13:58:51.0 & 26:06:24.0 &     9773 (3) &     3.5 (0.2) &    300 (0) &    285 (0) &     3.6 (0.2) &    296 (1) &    272 (1) &     3.6 (0.2) &    290 (4) &    278 (4) \\
     UGC  9007 & 14:05:06.0 & 09:20:21.0 &     4618 (4) &     1.4 (0.1) &     59 (6) &     54 (4) &     1.4 (0.1) &     66 (7) &     56 (1) &     1.4 (0.1) &     63 (4) &     57 (4) \\
     UGC  9134 & 14:16:35.2 & 09:59:09.0 &    11158 (19) &     4.1 (0.2) &    350 (30) &    334 (28) &     4.0 (0.2) &    359 (25) &    339 (12) &     4.0 (0.2) &    356 (21) &    337 (4) \\
     UGC  9450 & 14:39:45.4 & 23:23:50.0 &     4471 (9) &     2.4 (0.1) &    221 (13) &    205 (13) &     2.4 (0.1) &    223 (13) &    209 (10) &     2.4 (0.1) &    214 (5) &    207 (4) \\
     UGC  9535 & 14:48:42.5 & 12:27:25.0 &     1792 (1) &     8.6 (0.4) &    200 (5) &    192 (1) &     8.7 (0.4) &    205 (1) &    192 (1) &     8.7 (0.4) &    202 (4) &    194 (4) \\
     UGC  9698 & 15:05:27.2 & 23:41:18.0 &     4853 (6) &     2.7 (0.1) &    198 (21) &    172 (10) &     2.6 (0.1) &    207 (19) &    173 (20) &     2.7 (0.1) &    203 (9) &    177 (4) \\
     UGC  9803 & 15:17:17.8 & 29:24:00.0 &     5257 (2) &     1.5 (0.1) &     82 (1) &     72 (2) &     1.5 (0.1) &     80 (7) &     70 (5) &     1.6 (0.1) &     88 (7) &     73 (4) \\
     UGC  9901 & 15:34:27.0 & 12:16:12.0 &     3160 (1) &     5.9 (0.3) &    246 (3) &    237 (2) &     5.9 (0.3) &    247 (3) &    234 (3) &     5.9 (0.3) &    246 (4) &    236 (4) \\
UGC  9965\ddag & 15:40:06.6 & 20:40:50.0 &     4524 (11) &     4.7 (0.2) &    133 (20) &    113 (31) &     4.8 (0.2) &    153 (21) &    114 (36) &     4.7 (0.2) &    125 (14) &    118 (4) \\
     UGC  9979 & 15:42:19.3 & 00:28:31.0 &     1961 (3) &     4.8 (0.2) &    137 (4) &    126 (1) &     4.8 (0.2) &    144 (7) &    132 (3) &     4.8 (0.2) &    144 (4) &    129 (4) \\
     UGC 10042 & 15:48:57.2 & 07:13:18.0 &     4224 (5) &     4.6 (0.2) &    342 (5) &    335 (12) &     4.6 (0.2) &    350 (12) &    336 (1) &     4.6 (0.2) &    348 (4) &    340 (4) \\
     UGC 10243 & 16:10:29.6 & 19:57:15.0 &     7937 (2) &     4.6 (0.2) &    165 (5) &    146 (6) &     4.5 (0.2) &    162 (5) &    146 (5) &     4.6 (0.2) &    165 (4) &    147 (6) \\
UGC 10351\ddag & 16:21:28.2 & 28:38:25.0 &      891 (1) &     6.3 (0.3) &     70 (2) &     66 (0) &     6.4 (0.3) &     79 (1) &     65 (1) &     6.3 (0.3) &     71 (4) &     67 (4) \\
     UGC 10384 & 16:26:46.7 & 11:34:49.0 &     4966 (5) &     7.0 (0.4) &    384 (9) &    370 (3) &     7.0 (0.4) &    386 (22) &    373 (3) &     7.0 (0.3) &    383 (6) &    371 (4) \\
UGC 10628\ddag & 16:58:08.6 & 22:59:06.0 &     2980 (2) &     5.0 (0.3) &    259 (1) &    250 (3) &     5.1 (0.3) &    261 (7) &    247 (1) &     5.0 (0.3) &    260 (4) &    252 (4) \\
     UGC 10721 & 17:08:25.5 & 25:31:02.0 &     2920 (5) &     5.3 (0.3) &    286 (11) &    273 (12) &     5.3 (0.3) &    298 (17) &    276 (8) &     5.3 (0.3) &    292 (4) &    279 (4) \\
     UGC 10747 & 17:11:59.7 & 23:22:48.0 &     8821 (4) &     3.3 (0.2) &    317 (0) &    317 (0) &     3.3 (0.2) &    340 (1) &    313 (1) &     3.3 (0.2) &    327 (4) &    322 (4) \\
     UGC 10901 & 17:33:54.0 & 05:28:34.0 &     2834 (1) &     9.2 (0.5) &    190 (0) &    184 (0) &     9.2 (0.5) &    193 (1) &    185 (1) &     9.2 (0.5) &    195 (4) &    186 (4) \\
     UGC 11017 & 17:52:08.6 & 29:51:41.0 &     4644 (0) &     3.9 (0.2) &    183 (0) &    167 (0) &     3.9 (0.2) &    187 (1) &    173 (1) &     3.9 (0.2) &    186 (4) &    174 (4) \\
     UGC 11120 & 18:07:02.2 & 20:29:17.0 &     2242 (6) &     3.6 (0.2) &    179 (0) &    169 (10) &     3.6 (0.2) &    178 (13) &    167 (2) &     3.6 (0.2) &    182 (4) &    170 (4) \\
     UGC 11285 & 18:35:14.3 & 22:29:58.0 &     4501 (4) &     5.3 (0.3) &    306 (6) &    300 (8) &     5.4 (0.3) &    315 (9) &    303 (4) &     5.3 (0.3) &    314 (13) &    304 (4) \\
     UGC 11362 & 18:49:56.0 & 23:15:16.0 &     4205 (2) &     3.7 (0.2) &    197 (20) &    191 (11) &     3.7 (0.2) &    207 (14) &    192 (1) &     3.7 (0.2) &    204 (11) &    194 (4) \\
UGC 11461\ddag & 19:39:09.1 & 08:48:37.0 &     3122 (4) &     6.6 (0.3) &    322 (37) &    298 (37) &     6.6 (0.3) &    326 (10) &    299 (8) &     6.6 (0.3) &    324 (8) &    302 (4) \\
     UGC 11482 & 19:49:20.1 & 07:09:35.0 &     3160 (3) &     8.3 (0.4) &    289 (4) &    280 (8) &     8.4 (0.4) &    293 (7) &    278 (1) &     8.3 (0.4) &    290 (4) &    281 (4) \\
     UGC 11504 & 20:02:25.2 & 07:46:33.0 &     5620 (2) &     3.1 (0.2) &    150 (6) &    138 (5) &     3.1 (0.2) &    160 (7) &    142 (5) &     3.2 (0.2) &    163 (7) &    147 (4) \\
UGC 11578\ddag & 20:30:42.7 & 09:11:25.0 &     4601 (1) &     8.7 (0.4) &     99 (1) &     88 (1) &     8.7 (0.4) &    104 (2) &     88 (1) &     8.7 (0.4) &    100 (4) &     90 (4) \\
     UGC 11624 & 20:46:15.8 & 06:42:43.0 &     4831 (2) &     3.2 (0.2) &    218 (7) &    209 (8) &     3.2 (0.2) &    219 (4) &    206 (1) &     3.2 (0.2) &    219 (4) &    210 (4) \\
     UGC 11627 & 20:46:48.9 & 05:38:47.0 &     4864 (2) &     2.3 (0.1) &    126 (9) &    118 (6) &     2.3 (0.1) &    132 (10) &    117 (7) &     2.3 (0.1) &    134 (4) &    119 (4) \\
     UGC 11655 & 20:57:49.4 & 25:38:15.0 &     4754 (4) &     3.8 (0.2) &    266 (6) &    259 (9) &     3.8 (0.2) &    274 (10) &    262 (7) &     3.8 (0.2) &    275 (4) &    264 (4) \\
     UGC 11708 & 21:14:56.0 & 02:50:05.0 &     4164 (2) &     7.5 (0.4) &    292 (4) &    282 (7) &     7.4 (0.4) &    301 (4) &    283 (2) &     7.5 (0.4) &    296 (4) &    285 (4) \\
     UGC 11774 & 21:36:10.6 & 17:03:40.0 &     6875 (5) &     3.7 (0.2) &    235 (11) &    232 (10) &     3.7 (0.2) &    241 (9) &    232 (1) &     3.6 (0.2) &    238 (4) &    231 (4) \\
     UGC 11849 & 21:55:40.2 & 24:53:51.0 &     5840 (3) &     4.3 (0.2) &    357 (15) &    347 (12) &     4.3 (0.2) &    359 (9) &    348 (5) &     4.3 (0.2) &    358 (7) &    349 (4) \\
     UGC 11926 & 22:09:31.1 & 18:40:54.0 &     1652 (1) &     5.2 (0.3) &    110 (3) &    100 (1) &     5.2 (0.3) &    111 (4) &    100 (1) &     5.2 (0.3) &    112 (4) &    102 (4) \\
UGC 11992\ddag & 22:20:47.4 & 14:14:05.0 &     3592 (1) &     5.3 (0.3) &    174 (8) &    169 (2) &     5.3 (0.3) &    178 (1) &    169 (2) &     5.3 (0.3) &    179 (4) &    171 (4) \\
     UGC 12094 & 22:34:46.0 & 22:33:48.0 &     7607 (9) &     1.8 (0.1) &     59 (16) &     59 (32) &     1.7 (0.1) &     86 (13) &     55 (4) &     1.8 (0.1) &     66 (16) &     71 (4) \\
     UGC 12158 & 22:42:10.5 & 19:59:49.0 &     9289 (5) &     4.3 (0.2) &    178 (7) &    160 (7) &     4.3 (0.2) &    187 (6) &    164 (4) &     4.3 (0.2) &    184 (4) &    164 (4) \\
     UGC 12388 & 23:08:30.2 & 12:49:49.0 &     4584 (3) &     6.3 (0.3) &    272 (14) &    266 (12) &     6.2 (0.3) &    281 (10) &    266 (3) &     6.3 (0.3) &    277 (6) &    269 (4) \\
   \obc\ P02-3 & 23:16:59.0 & 07:52:19.0 &     3178 (3) &     1.9 (0.1) &    109 (5) &    103 (7) &     1.8 (0.1) &    108 (5) &     99 (2) &     1.9 (0.1) &    112 (6) &    103 (4) \\
     UGC 12537 & 23:21:07.9 & 29:32:58.0 &     6112 (6) &     3.6 (0.2) &    445 (16) &    445 (12) &     3.5 (0.2) &    452 (19) &    448 (11) &     3.6 (0.2) &    452 (5) &    448 (4) \\
   \obc\ P07-1 & 23:22:58.5 & 07:40:20.0 &     3472 (6) &     1.4 (0.1) &    171 (20) &    162 (24) &     1.4 (0.1) &    171 (10) &    165 (6) &     1.4 (0.1) &    176 (6) &    168 (6) \\
     UGC 12624 & 23:29:07.6 & 21:33:51.0 &     3516 (2) &     3.5 (0.2) &    194 (1) &    183 (3) &     3.5 (0.2) &    201 (4) &    186 (2) &     3.5 (0.2) &    198 (7) &    187 (4) \\
     UGC 12695 & 23:36:02.2 & 12:52:32.0 &     6185 (2) &     4.2 (0.2) &     85 (8) &     78 (4) &     4.2 (0.2) &     95 (11) &     78 (4) &     4.2 (0.2) &     89 (5) &     80 (4) \\
     UGC 12910 & 00:01:28.3 & 05:23:22.0 &     3948 (4) &     2.8 (0.1) &     74 (9) &     63 (3) &     2.8 (0.1) &     83 (4) &     66 (4) &     2.8 (0.1) &     77 (4) &     67 (4) \\
\enddata
\tablecomments
{\dag This galaxy may have \HI\ extending beyond
the Arecibo beam, resulting in the flux measurements reported here
being too low.  No GBT follow-up results are availale.
See Section~\ref{sec:flux} for more details.\\
\ddag This galaxy has \HI\ extending beyond
the Arecibo beam, resulting in the flux measurements reported here
being too low.  The results from the GBT observations
(Table~\ref{tab:gbt}) are more reliable.  See Section~\ref{sec:flux} for more details.}
\end{deluxetable}

\begin{deluxetable}{lccccc}
\tablewidth{0pt}
\tablecaption{Results from GBT observatons \label{tab:gbt}}
\tablehead{
\colhead{Name} & \colhead{V$_{HEL}$} &
\colhead{Flux} & \colhead{rms} & \colhead{W$_{20}$} & \colhead{W$_{50}$} \\
   & \colhead{[\kms]} &\colhead{[Jy \kms]} & 
\colhead{[Jy]} & \colhead{[\kms]} & \colhead{[\kms]} \\
}
\startdata
            UGC   941&    2729 (8) &     3.7 (0.4) &     3.3 &      95 (16) &      74 (16) \\
            UGC  2885&    5799 (8) &    20.7 (2.1) &     4.5 &     593 (16) &     576 (16) \\
            UGC  2922&    5210 (8) &     5.0 (0.5) &     3.6 &     411 (16) &     390 (16) \\
        UGC  3118\dag&    8297 (8) &     1.8 (0.2) &     2.9 &      86 (16) &      78 (16) \\
            UGC  5160&    6653 (8) &     2.9 (0.3) &     1.2 &     405 (16) &     387 (16) \\
            UGC  5215&    5475 (8) &     7.5 (0.8) &     1.5 &     440 (16) &     420 (16) \\
            UGC  5218&    3147 (8) &     7.7 (0.8) &     3.0 &     260 (16) &     192 (16) \\
            UGC  6886&    6984 (8) &     4.8 (0.5) &     1.9 &     388 (16) &     371 (16) \\
            UGC  8091&     213 (8) &     8.8 (0.9) &     3.5 &      53 (16) &      41 (16) \\
            UGC  8249&    2540 (8) &     9.0 (0.9) &     3.1 &     191 (16) &     181 (16) \\
            UGC  8503&    4674 (8) &     2.3 (0.3) &     8.3 &      85 (16) &      78 (16) \\
            UGC  9007&    4617 (8) &     1.5 (0.2) &     2.9 &      63 (16) &      52 (16) \\
            UGC  9450&    4487 (8) &     2.6 (0.3) &     2.2 &     206 (16) &     197 (16) \\
            UGC  9803&    5259 (8) &     1.5 (0.2) &     2.7 &      78 (16) &      69 (16) \\
            UGC  9965&    4526 (8) &     5.3 (0.5) &     3.1 &     126 (16) &     101 (16) \\
            UGC 10351&     891 (8) &     9.0 (0.9) &     2.3 &      78 (16) &      63 (16) \\
            UGC 10628&    2979 (8) &     5.8 (0.6) &     2.1 &     273 (16) &     247 (16) \\
            UGC 11017&    4648 (8) &     4.4 (0.4) &     2.2 &     197 (16) &     182 (16) \\
            UGC 11461&    3125 (8) &     8.1 (0.8) &     2.5 &     319 (16) &     303 (16) \\
            UGC 11578&    4602 (8) &    10.9 (1.1) &     3.0 &     103 (16) &      86 (16) \\
            UGC 11627&    4861 (8) &     2.4 (0.2) &     2.1 &     114 (16) &     113 (16) \\
            UGC 11992&    3597 (8) &     6.2 (0.6) &     2.4 &     185 (16) &     172 (16) \\
        UGC 12094\dag&    7589 (8) &     1.6 (0.2) &     3.6 &      56 (16) &      44 (16) \\
\enddata
\tablecomments{\dag The heliocentric velocity found by the
GBT for this galaxy is off by $\sim$17 \kms\ due to an instrumental error.}
\end{deluxetable}

\begin{deluxetable}{llllllll}
\tabletypesize{\scriptsize}
\tablewidth{0pt}
\tablecaption{Literature Values \label{tab:lit}}
\tablehead{
\colhead{Name} & \colhead{Flux} & \colhead{W$_{20}$} &
\colhead{W$_{50}$} & \colhead{v$_{HEL}$} &
\colhead{Tel.$^1$} & \colhead{Refs.}\\
& \colhead{[Jy \kms]} & \colhead{[\kms]} &
\colhead{[\kms]} & \colhead{[\kms]}
}
\startdata
     UGC   199 &       3.13 ( 0.2) & 122 (3) &        106 (2) &       1803 (4) & A &  a\\
     NGC   257 &       12.5 ( 1.2$^2$) & 426 (4) &        402 (4) &       5278 (5) & A &  b\\
     NGC   257 &       11.7 ( 1.2$^2$) & 437 (4) &        394 (4) &       5270 (5) & A &  c\\
     UGC   317 &      0.970 ( 0.3) & 47 (7) &         30 (4) &       5264 (5) & A &  a\\
     UGC   317 &      0.990 ( 0.1$^2$) & 41 (9) &         27 (6) &       5361 (5) & G &  d\\
     UGC   628 &       4.52 ( 0.5$^2$) & 239 (8$^3$) &        222 (8$^3$) &       5447 (8$^3$) & A &  e\\
     UGC   628 &       4.73 ( 0.3) & 251 (6) &        229 (4) &       5446 (4) & A &  a\\
     UGC   628 &       4.20 ( 1.1) & 237 (10) &        224 (7) &       5443 (5) & G &  d\\
     UGC   941 &       2.02 ( 0.2) & 88 (8) &         47 (5) &       2726 (5) & A &  a\\
     UGC   989 &       5.71 ( 0.3) & 166 (4) &        146 (2) &       5107 (5) & A &  a\\
     UGC  1085 &       1.68 ( 0.2) & 58 (3) &         38 (2) &        652 (4) & A &  a\\
     UGC  1246 &       7.35 ( 0.2) & 109 (2) &         82 (2) &        804 (4) & A &  a\\
     UGC  1519 &       2.27 ( 0.2) & 132 (4) &        117 (3) &       2346 (4) & A &  a\\
     UGC  1519 &       2.13 ( 0.2$^2$) & 123 (8$^3$) &        111 (8$^3$) &       2346 (8$^3$) & A &  e\\
     UGC  2008 &       1.93 ( 0.3) & 124 (5) &        109 (8) &       5041 (5) & A &  a\\
     UGC  2008 &       2.69 ( 0.5) & 147 (15$^3$) &        127 (15$^3$) &       5042 (15$^3$) & A &  f\\
     UGC  2339 &       2.17 ( 0.2$^2$) & 334 (15$^3$) &        325 (15$^3$) &       4821 (15$^3$) & A &  f\\
     UGC  2339 &       4.46 ( 0.4$^2$) & 478 (8$^3$) &        457 (8$^3$) &      16469 (8$^3$) & A &  e\\
     UGC  2339 &       4.33 ( 0.4$^2$) & 457 (8$^3$) &    \nodata  &      16469 (8$^3$) & A &  g\\
     UGC  2432 &       5.67 ( 0.3) & 107 (3) &         88 (2) &        764 (4$^3$) & A &  a\\
     UGC  2602 &       7.01 ( 0.7$^2$) & 298 (8$^3$) &        276 (8$^3$) &      10099 (8$^3$) & A &  e\\
     UGC  2602 &       5.10 ( 0.8) & 305 (25) &        267 (17) &      10098 (8) & N &  h\\
     UGC  2809 &       9.29 ( 0.2) & 155 (2) &        131 (1) &       1293 (4) & A &  a\\
     UGC  2885 &       25.2 ( 2.5$^2$) & 629 (34$^3$) &        583 (34$^3$) &       5794 (34$^3$) & W &  i\\
     UGC  2885 &       34.2 ( 3.4) & 595 (4) &        550 (4) &       5801 (4) & J &  j\\
     UGC  2885 &       17.3 ( 1.7$^2$) & 581 (4) &        556 (4) &       5802 (15$^3$) & A &  k\\
     UGC  2885 &       32.7 ( 3.3$^2$) & 607 (4) &    \nodata  &       5801 (2) & G &  l\\
     UGC  2905 &       4.67 ( 0.5$^2$) & 71 (8$^3$) &         47 (8$^3$) &        293 (8$^3$) & A &  e\\
     UGC  2905 &       4.47 ( 0.1) & 71 (1) &         48 (1) &        292 (4) & A &  a\\
     UGC  2922 &       5.12 ( 0.5$^2$) & 370 (8$^3$) &        351 (8$^3$) &       5215 (8$^3$) & A &  e\\
     UGC  2975 &       2.41 ( 0.2$^2$) & 312 (3) &        280 (3) &      19937 (8$^3$) & A &  m\\
     UGC  3044 &       2.20 ( 0.3) & 246 (9) &        228 (5) &       5234 (5) & A &  a\\
     UGC  3118 &       1.72 ( 0.2) & 89 (8) &         55 (5) &       8310 (5) & A &  a\\
     UGC  3118 &      0.620 ( 0.4) & 65 (13) &         51 (9) &       8316 (6) & G &  d\\
     UGC  3169 &       2.79 ( 0.3$^2$) & 373 (3) &        358 (3) &       8788 (3) & A &  n\\
     UGC  3247 &       5.99 ( 0.6$^2$) & 196 (2) &        137 (3) &       3369 (3) & A &  n\\
     UGC  3247 &       5.15 ( 0.5$^2$) & 213 (4) &        143 (4) &       3373 (4) & A &  o\\
     UGC  3296 &       10.5 ( 1.1$^2$) & 335 (3) &        315 (3) &       4269 (3) & A &  p\\
     UGC  3296 &       7.56 ( 0.8$^2$) & 348 (4) &        318 (4) &       4264 (15$^3$) & A &  k\\
     UGC  3330 &       5.22 ( 0.5$^2$) & 204 (3) &        163 (3) &       5234 (3) & A &  n\\
     UGC  3330 &       4.52 ( 0.5$^2$) & 202 (4) &        170 (4) &       5239 (4) & A &  o\\
     UGC  3476 &       12.3 ( 0.6) & 141 (9) &         83 (5) &        469 (5) & G &  d\\
     UGC  3476 &       7.33 ( 0.3) & 117 (6) &         70 (4) &        469 (4) & A &  a\\
     UGC  3564 &       6.26 ( 0.6$^2$) & 94 (3) &         79 (3) &       2548 (3) & A &  n\\
     UGC  3564 &       5.47 ( 0.2) & 97 (2) &         75 (1) &       2548 (4) & A &  a\\
     UGC  3582 &       2.94 ( 0.3$^2$) & 171 (3) &        152 (3) &        839 (3) & A &  n\\
     UGC  3621 &       9.56 ( 0.5) & 177 (4) &        159 (2) &       2337 (4) & A &  a\\
     UGC  3621 &       9.59 ( 1.0$^2$) & 173 (3) &        158 (3) &       2337 (3) & A &  n\\
     UGC  3621 &       10.3 ( 1.0$^2$) & 172 (4) &        156 (4) &       2336 (4) & A &  o\\
     UGC  3755 &       8.36 ( 0.8$^2$) & 84 (3) &         43 (3) &        315 (3) & A &  n\\
     UGC  3755 &       8.49 ( 0.3) & 70 (2) &         37 (1) &        314 (4) & A &  a\\
     UGC  3767 &       4.49 ( 0.4$^2$) & 264 (3) &        246 (3) &       5809 (3) & A &  n\\
     UGC  3767 &       3.38 ( 0.3) & 273 (7) &        254 (5) &       5806 (5) & A &  a\\
     UGC  3946 &       11.2 ( 0.3) & 108 (3) &         76 (2) &       1195 (4) & A &  a\\
     UGC  3946 &       11.8 ( 1.2$^2$) & 112 (3) &         79 (3) &       1194 (3) & A &  n\\
     UGC  3946 &       14.3 ( 1.4$^2$) & 126 (4) &    \nodata  &       1199 (1) & G &  l\\
     UGC  4117 &       3.55 ( 0.4) & 80 (6) &         59 (4) &        775 (4) & A &  a\\
     UGC  4180 &       6.16 ( 0.6$^2$) & 343 (8$^3$) &        330 (8$^3$) &       5225 (8$^3$) & A &  g\\
     UGC  4180 &       5.18 ( 0.5) & \nodata  &        347 (4$^3$) &       5223 (4$^3$) & A &  q\\
     UGC  4180 &       5.50 ( 0.6$^2$) & 358 (4) &        333 (4) &       5231 (4) & A &  k\\
     UGC  4385 &       6.60 ( 0.7$^2$) & \nodata  &        171 (5) &       1969 (8$^3$) & A &  r\\
     UGC  4660 &       6.23 ( 0.6$^2$) & 88 (3) &         61 (3) &       2203 (3) & A &  n\\
     UGC  4660 &       5.67 ( 0.6$^2$) & 69 (4) &         57 (4) &       2201 (4) & A &  r\\
     UGC  4660 &       7.30 ( 0.2) & 75 (4) &    \nodata  &       2205 (2) & G &  l\\
     UGC  4780 &       7.54 ( 0.2) & 167 (4) &        128 (2) &       3292 (4) & A &  a\\
     UGC  4780 &       9.13 ( 0.9$^2$) & 147 (3) &        124 (3) &       3283 (3) & A &  n\\
     UGC  4955 &       1.98 ( 0.2) & 225 (12) &        199 (8) &       6462 (8) & A &  a\\
     UGC  5030 &       2.94 ( 0.3$^2$) & 300 (3) &        286 (3) &       2148 (\nodata) & A &  n\\
     UGC  5030 &       3.72 ( 0.4$^2$) & 307 (4) &        285 (4) &       2150 (15$^3$) & A &  k\\
     UGC  5160 &       4.27 ( 0.4$^2$) & 416 (4) &        381 (4) &       6663 (15$^3$) & A &  k\\
     UGC  5215 &       9.05 ( 0.9$^2$) & 427 (4) &        404 (4) &       5484 (15$^3$) & A &  k\\
     UGC  5218 &       7.01 ( 0.7$^2$) & 194 (3) &        183 (3) &       3085 (3) & A &  n\\
     UGC  5218 &       12.7 ( 1.3$^2$) & 195 (4) &        182 (4) &       3089 (15$^3$) & A &  k\\
     UGC  5326 &       3.40 ( 0.3) & 118 (7) &         92 (4) &       1417 (5) & A &  a\\
     UGC  5326 &       4.04 ( 0.4$^2$) & 124 (4) &        101 (4) &       1411 (15$^3$) & A &  k\\
     UGC  5326 &       3.65 ( 0.4$^2$) & 126 (4) &        101 (4) &       1413 (4) & A &  s\\
     UGC  5358 &       4.72 ( 0.5$^2$) & 217 (4) &        205 (4) &       2914 (15$^3$) & A &  k\\
     UGC  5629 &       2.89 ( 0.1) & 133 (3) &        114 (6) &       6203 (5) & A &  a\\
     UGC  5651 &       8.33 ( 0.8$^2$) & 229 (4) &        213 (4) &       5568 (15$^3$) & A &  k\\
     UGC  5852 &       10.9 ( 1.1$^2$) & 368 (5) &    \nodata  &       6184 (10$^3$) & G &  t\\
     UGC  5852 &       13.1 ( 1.3$^2$) & 378 (22$^3$) &    \nodata  &       6165 (22$^3$) & A &  u\\
     UGC  5852 &       12.4 ( 1.2$^2$) & 379 (4) &        338 (4) &       6177 (4) & A &  n\\
     UGC  6018 &       1.69 ( 0.2) & 72 (3) &         51 (3) &       1292 (4$^3$) & A &  n\\
     UGC  6018 &       1.76 ( 0.4) & 65 (7) &         48 (5) &       1295 (5) & A &  a\\
     UGC  6421 &       7.31 ( 0.3) & 165 (3) &        165 (3) &       1595 (3) & A &  n\\
     UGC  6421 &       6.59 ( 0.2) & 179 (2) &        163 (1) &       1504 (4) & A &  a\\
     UGC  6464 &       3.97 ( 0.4$^2$) & 163 (4) &        145 (4) &       1071 (15$^3$) & A &  b\\
     UGC  6464 &       2.60 ( 0.3$^2$) & \nodata  &        125 (8$^3$) &       1067 (8$^3$) & A &  v\\
     UGC  6464 &       3.11 ( 0.3$^2$) & 150 (4) &        128 (4) &       1080 (15$^3$) & A &  k\\
     UGC  6476 &       3.79 ( 0.4$^2$) & 300 (4) &        322 (4) &       7328 (15$^3$) & A &  k\\
     UGC  6872 &       6.32 ( 0.6$^2$) & 216 (4) &        183 (4) &       3205 (15$^3$) & A &  k\\
     UGC  6872 &       6.52 ( 0.7$^2$) & 263 (4) &        193 (4) &       3212 (8) & A &  w\\
     UGC  6886 &       5.54 ( 0.6$^2$) & 366 (4) &        352 (4) &       6980 (15$^3$) & A &  k\\
     UGC  6886 &       5.10 ( 0.5$^2$) & 394 (4) &        360 (4) &       6980 (10$^3$) & A & no\\
     UGC  6886 &       7.48 ( 0.7$^2$) & 406 (4) &        350 (4) &       6972 (8$^3$) & A &  x\\
     UGC  7302 &       3.87 ( 0.3) & 93 (4) &         74 (2) &       3838 (4) & A &  a\\
     UGC  7302 &       3.84 ( 0.4$^2$) & 93 (4$^3$) &         74 (2) &       3836 (4) & A &  y\\
     UGC  7666 &       2.54 ( 0.2) & 138 (10) &        101 (6) &       2559 (5) & A &  a\\
     UGC  7976 &       6.50 ( 0.7) & 98 (4) &         75 (4) &       2649 (5) & E &  z\\
     UGC  7976 &       5.34 ( 0.1) & 93 (2) &         73 (1) &       2668 (4) & A &  a\\
     UGC  8091 &       8.70 ( 0.1) & 48 (4) &    \nodata  &        214 (1) & G &  l\\
     UGC  8091 &       7.76 ( 0.8$^2$) & 41 (4) &         27 (4) &        213 (15$^3$) & A &  k\\
     UGC  8091 &       8.60 ( 1.3) & 45 (4) &    \nodata  &        212 (16$^3$) & B & aa\\
     UGC  8091 &       5.91 ( 0.6$^2$) & 46 (4) &         30 (4) &        214 (\nodata) & A & bb\\
     UGC  8091 &       6.20 ( 0.6$^2$) & 40 (4) &         26 (4) &        214 (10$^3$) & A & cc\\
     UGC  8091 &       8.40 ( 1.1) & 38 (4) &         24 (4) &        214 (22$^3$) & E &  z\\
     UGC  8249 &       6.95 ( 0.3) & 194 (3) &        175 (2) &       2541 (4) & A &  a\\
     UGC  8249 &       9.00 ( 0.4) & 195 (4) &    \nodata  &       2545 (\nodata) & G &  l\\
     UGC  8503 &       1.81 ( 0.2) & 71 (6) &         40 (4) &       4666 (4) & A &  a\\
     UGC  8516 &       4.04 ( 0.4) & 146 (5) &    \nodata  &       1011 (15) & G &  t\\
     UGC  8516 &       4.00 ( 0.3) & 125 (5) &    \nodata  &       1021 (4$^3$) & A &  r\\
     UGC  8516 &       3.88 ( 0.4$^2$) & 124 (4) &        101 (4) &       1025 (15$^3$) & A &  k\\
     UGC  8516 &       4.19 ( 0.4$^2$) & 126 (4) &         99 (4) &       1023 (2) & A & dd\\
     UGC  8896 &       5.87 ( 0.6$^2$) & 304 (4) &        290 (4) &       4401 (15$^3$) & A &  k\\
     UGC  9007 &       1.42 ( 0.1$^2$) & 55 (4) &         40 (4) &       4619 (15$^3$) & A &  k\\
     UGC  9007 &       1.60 ( 0.1) & 57 (4) &    \nodata  &       4625 (\nodata) & G &  l\\
     UGC  9007 &       1.20 ( 0.1$^2$) & 148 (4) &         62 (11) &       4585 (11) & A & ee\\
     UGC  9134 &       4.65 ( 1.0) & 332 (3) &    \nodata  &      11157 (15$^3$) & A & ff\\
     UGC  9450 &       3.05 ( 1.1) & 222 (3) &    \nodata  &       4462 (8) & A & ff\\
     UGC  9450 &       2.10 ( 0.2) & 213 (12) &        180 (7) &       4481 (5) & A &  a\\
     UGC  9450 &       3.18 ( 0.9) & 201 (6) &        192 (4) &       4467 (4) & G &  d\\
     UGC  9535 &       8.61 ( 0.4) & 193 (4) &        183 (4) &       1792 (15$^3$) & A &  k\\
     UGC  9535 &       7.99 ( 2.3) & 197 (3) &    \nodata  &        837 (8) & A & ff\\
     UGC  9698 &       2.86 ( 1.3) & 180 (3) &    \nodata  &       4852 (8) & A & ff\\
     UGC  9803 &       1.69 ( 1.0) & 90 (3) &    \nodata  &       5259 (8) & A & ff\\
     UGC  9803 &       2.44 ( 0.6) & 82 (17) &         45 (11) &       5267 (7) & G &  d\\
     UGC  9803 &       1.49 ( 0.1) & 81 (5) &         61 (3) &       5256 (4) & A &  a\\
     UGC  9901 &       5.79 ( 2.0) & 241 (4) &        223 (4) &       3161 (15$^3$) & A &  k\\
     UGC  9901 &       5.27 ( 0.5$^2$) & 248 (3) &    \nodata  &       3166 (8) & A & ff\\
     UGC  9965 &       2.40 ( 0.4) & 115 (4) &         80 (4) &       4525 (8) & A & no\\
     UGC  9965 &       5.37 ( 0.5$^2$) & 115 (4) &         88 (4) &       4528 (15$^3$) & A &  k\\
     UGC  9979 &       5.44 ( 0.5$^2$) & 134 (4) &        118 (4) &       1961 (15$^3$) & A &  k\\
     UGC  9979 &       5.20 ( 1.1) & 140 (4) &    \nodata  &       1961 (10) & G &  l\\
     UGC 10042 &       4.71 ( 0.5$^2$) & 349 (4) &        337 (4) &       4226 (15$^3$) & A &  k\\
     UGC 10243 &       4.56 ( 1.9) & 161 (3) &    \nodata  &       7937 (8) & A & ff\\
     UGC 10351 &       5.70 ( 0.4) & 67 (3) &         43 (2) &        892 (4) & A &  a\\
     UGC 10351 &       9.45 ( 0.4) & 77 (11$^3$) &         48 (2) &        891 (4) & G &  d\\
     UGC 10384 &       5.74 ( 2.3) & 382 (3) &    \nodata  &       4970 (8) & A & ff\\
     UGC 10628 &       6.13 ( 0.5) & 272 (5) &    \nodata  &       2974 (15$^3$) & G &  t\\
     UGC 10628 &       5.99 ( 0.6) & 259 (4) &        240 (4) &       2980 (15$^3$) & A &  k\\
     UGC 10628 &       6.28 ( 0.2) & 259 (6) &        244 (5) &       6267 (5) & A &  b\\
     UGC 10721 &       6.10 ( 0.6$^2$) & 286 (4) &        267 (4) &       2918 (15$^3$) & A &  k\\
     UGC 10747 &       3.70 ( 0.4$^2$) & 330 (4) &        299 (4) &       8820 (15$^3$) & A &  k\\
     UGC 10901 &       8.33 ( 0.3) & 198 (3) &        178 (2) &       2834 (4) & A &  a\\
     UGC 11017 &       3.74 ( 0.3) & 189 (6) &        167 (4) &       4648 (4) & A &  a\\
     UGC 11017 &       1.40 ( 0.4) & 149 (4) &        143 (4) &       4628 (13) & A & no\\
     UGC 11120 &       3.58 ( 0.2) & 179 (3) &        162 (2) &       2242 (4) & A &  a\\
     UGC 11285 &       5.01 ( 0.2) & 321 (5) &        299 (3) &       4500 (4) & A &  a\\
     UGC 11362 &       3.74 ( 0.2) & 207 (4) &        190 (3) &       4204 (4) & A &  a\\
     UGC 11461 &       8.10 ( 0.9) & 302 (12) &        286 (8) &       3121 (4) & N &  h\\
     UGC 11482 &       8.21 ( 0.3) & 289 (2) &        273 (1) &       3163 (4) & A &  a\\
     UGC 11482 &       9.04 ( 1.5) & 271 (6) &        287 (10) &       3157 (5) & G &  d\\
     UGC 11504 &       3.50 ( 0.4) & 168 (12) &        128 (19) &       5619 (7) & A &  a\\
     UGC 11578 &       8.69 ( 0.3) & 96 (2) &         73 (2) &       4604 (4) & A &  a\\
     UGC 11578 &       11.8 ( 0.7) & 103 (5) &         77 (3) &       4602 (4) & G &  d\\
     UGC 11624 &       2.91 ( 0.2) & 218 (4) &        201 (3) &       4831 (4) & A &  a\\
     UGC 11627 &       2.14 ( 0.2) & 135 (9) &        114 (6) &       4864 (5) & A &  a\\
     UGC 11627 &       1.18 ( 0.4) & 117 (19) &        195 (7) &       4866 (5) & G &  d\\
     UGC 11655 &       3.81 ( 0.6) & 273 (17) &        247 (11) &       4761 (7) & G &  d\\
     UGC 11655 &       3.26 ( 0.8) & 277 (9) &        255 (6) &       4756 (5) & A &  a\\
     UGC 11655 &       4.19 ( 0.7) & 269 (3) &        259 (3) &       4756 (3) & A & gg\\
     UGC 11708 &       8.39 ( 0.8$^2$) & 302 (4) &        283 (4) &       4161 (4) & A &  k\\
     UGC 11774 &       3.48 ( 0.3) & 255 (9) &        228 (6) &       6878 (5) & A &  a\\
     UGC 11774 &       3.92 ( 0.4$^2$) & 245 (8$^3$) &        228 (8$^3$) &       7126 (8$^3$) & A &  e\\
     UGC 11849 &       5.09 ( 0.8) & 361 (3) &        339 (3) &       5841 (1) & A & gg\\
     UGC 11849 &       3.60 ( 0.8) & 337 (25) &        320 (17) &       5834 (8) & N &  h\\
     UGC 11926 &       5.29 ( 0.5$^2$) & 111 (8$^3$) &         99 (8$^3$) &       1649 (8$^3$) & A &  e\\
     UGC 11926 &       6.19 ( 0.3) & 115 (3) &         97 (2) &       1651 (4) & A &  a\\
     UGC 11992 &       5.43 ( 0.3) & 184 (3) &        167 (2) &       3596 (4) & A &  a\\
     UGC 11992 &       7.12 ( 1.1) & 181 (11) &        162 (7) &       3596 (5) & G &  d\\
     UGC 12094 &       2.27 ( 0.2) & 73 (6) &         41 (4) &       7614 (4) & A &  a\\
     UGC 12094 &       1.14 ( 0.2) & 62 (3) &         32 (3) &       7604 (3) & A & gg\\
     UGC 12158 &       4.87 ( 0.5$^2$) & 178 (8$^3$) &        156 (8$^3$) &       9290 (8$^3$) & A &  e\\
     UGC 12388 &       5.42 ( 0.2) & 281 (4) &        261 (3) &       4585 (4) & A &  a\\
     UGC 12537 &       3.81 ( 0.6) & 466 (3) &        441 (3) &       6116 (2) & A &  f\\
     UGC 12624 &       3.05 ( 0.3$^2$) & 184 (8$^3$) &        164 (8$^3$) &       3519 (8$^3$) & A &  e\\
     UGC 12624 &       3.19 ( 0.3) & 211 (11) &        183 (7) &       3517 (5) & A &  a\\
     UGC 12695 &       6.06 ( 0.6) & 87 (6) &         64 (4) &       6184 (4) & A &  a\\
     UGC 12695 &       2.80 ( 0.8) & 79 (18) &         63 (12) &       6189 (6) & N &  h\\
     UGC 12695 &       5.16 ( 0.5$^2$) & 92 (4) &         67 (4) &       6184 (4) & A &  o\\
     UGC 12910 &       3.36 ( 0.3) & 86 (7) &         53 (4) &       6352 (4) & A &  a\\
   CAP 0718+06 &       1.50 ( 0.1) & \nodata  &        199 (4) &       1141 (4) & A & hh\\
       Malin 1 &       3.50 ( 0.5) & 341 (4) &    \nodata  &      24750 (9$^3$) & A & ii\\
       Malin 1 &       4.60 ( 0.5$^2$) & 340 (4) &        315 (4) &      24745 (10$^3$) & B & jj\\
       Malin 1 &       2.70 ( 0.3$^2$) & 355 (4) &        295 (4) &      24705 (10$^3$) & A & jj\\
   \obc\ P02-3 &       1.82 ( 0.2) & 112 (5) &         95 (5) &       3178 (5) & A & kk\\
   \obc\ P07-1 &       1.43 ( 0.1) & 187 (5) &        177 (5) &       7949 (5) & A & kk\\
   \obc\ U01-4 &       1.93 ( 0.2) & 191 (5) &        181 (5) &       3450 (5) & A & kk\\
\enddata
\tablecomments{$^1$Telescopes:
 A= Arecibo 305m telescope;
 G = Green Bank 300ft telescope;
 N = Nancay Decimetric Radio Telescope;
 W = Westerbork Synthesis Radio Telescope;
 J = Jodrell Bank 250ft Telescope;
 E = Effelsberg 100m telescope;
 B = Green Bank 140ft Telescope.\\
$^2$ Not given in the literature, and estimated from assuming a 10\% flux error.\\
$^3$ Not given in the literature, and estimated from 0.5*channel resolution.}
\tablerefs{
a. \citet{1990ApJS...72..245};
b. \citet{1985ApJS...58..623};
c. \citet{1984ApJ...278..475};
d. \citet{1992ApJS...81....5};
e. \citet{1993AJ....105.1271};
f. \citet{1985AJ.....90.2445};
g. \citet{1997AJ....113.1197};
h. \citet{1998AAS..130..333};
i. \citet{1985AA...146..213};
j. \citet{1987MNRAS.224..953};
k. \citet{1985ApJS...59..161};
l. \citet{1988ApJS...67....1};
m. \citet{1989AJ.....97..633};
n. \citet{1986AJ.....91..705};
o. \citet{1987ApJS...63..515};
p. \citet{1986AJ.....91..732};
q. \citet{1985ApJ...292..404};
r. \citet{1984AJ.....89..758};
s. \citet{1986AJ.....92..742};
t. \citet{1979ApJS...40..527};
u. \citet{1983ApJS...53..269};
v. \citet{1982AJ.....87.1443};
w. \citet{1989HRHI.C...0000};
x. \citet{1989ApJS...69...65};
y. \citet{1984AA...138...85};
z. \citet{1986AAS...64..111};
aa. \citet{1985AJ.....90.1789};
bb. \citet{1985ApJ...289L..15};
cc. \citet{1984ApJ...285..453};
dd. \citet{1993AAS...99..379};
ee. \citet{1985ApJS...57..423};
ff. \citet{1988AJ.....96.1791};
gg. \citet{1986AJ.....92..250};
hh. \citet{1997AJ....113..905};
ii. \citet{1986ApJ...308..510};
jj. \citet{1989ApJ...341...89};
kk. \citet{2000AJ....119..136};
ll. \citet{1991AAS...87..425}.
}
\end{deluxetable}

\twocolumn




\end{document}